\documentclass[manuscript, screen]{acmart}

\AtBeginDocument{%
  }

\setcopyright{acmlicensed}
\copyrightyear{2026}
\acmYear{2026}
\acmDOI{XXXXXXX.XXXXXXX}
\acmConference[Conference CHI '26]{CHI conference on Human Factors in Computing Systems}{June 03--05, 2026}{Barcelona, Spain}
\acmISBN{978-1-4503-XXXX-X/2018/06}

\usepackage[utf8]{inputenc}

\usepackage{amsmath,amssymb}
\DeclareUnicodeCharacter{2265}{\ensuremath{\geq}}
\usepackage{newunicodechar}
\newunicodechar{≤}{\ensuremath{\leq}}

\usepackage{booktabs}
\usepackage{tabularx}   
\usepackage{array}      
\usepackage{ragged2e}   
\usepackage{threeparttable} 

\newcolumntype{L}{>{\RaggedRight\arraybackslash}X}

\usepackage{multirow}

\usepackage{enumitem}

\usepackage{hyperref}

\usepackage{soul}   

\usepackage[normalem]{ulem}  

\usepackage{tikz} 
\newcommand*\circled[1]{\tikz[baseline=(char.base)]{
            \node[shape=circle,draw,inner sep=0.5pt] (char) {#1};}} 
\usepackage[utf8]{inputenc}
\usepackage[T1]{fontenc}
\usepackage{textcomp}
\usepackage[most]{tcolorbox} 
\usepackage{xcolor}
\tcbuselibrary{breakable,listings} 

\usepackage{amssymb}

\definecolor{stringcolor}{RGB}{42,0,255}
\definecolor{keywordcolor}{RGB}{0,0,0}

\lstdefinelanguage{JSON}{
    keywords={true,false,null},
    keywordstyle=\color{keywordcolor}\bfseries,
    string=[s]{"}{"},
    stringstyle=\color{stringcolor},
    comment=[l]{//},
    commentstyle=\color{gray},
    morecomment=[s]{/*}{*/},
    morestring=[b]',
    alsoletter={:},
    alsodigit={-}
}

\newtcolorbox{mambajsonbox}{
  breakable,
  colback=white,
  colframe=black,
  boxrule=0.5pt,
  arc=0mm,
  parbox=true,
  title=prompt,
  left=3mm, right=3mm, top=2mm, bottom=3mm
}

\lstdefinestyle{embeddedStyle}{
    language=JSON,
    basicstyle=\ttfamily\small,
    numbers=left,
    numberstyle=\tiny\color{gray},
    stringstyle=\color{blue},
    keywordstyle=\color{purple},
    breaklines=true,
    frame=none, 
    backgroundcolor=\color{white}, 
    xleftmargin=1.5em, 
    showstringspaces=false,  
    postbreak={},            
    escapeinside={||}{||},  
    literate={_}{{{\char`\_}}}1, 
    literate={-}{{{\char`\-}}}1,
    literate={μ}{{\textmu}}1,       
    literate={°}{{\textdegree}}1    
}
\lstdefinestyle{plainTextStyle}{
    basicstyle=\ttfamily\small, 
    breaklines=true,             
    literate={_}{{{\char`\_}}}1, 
    literate={-}{{{\char`\-}}}1,
    literate={μ}{{\textmu}}1,       
    literate={°}{{\textdegree}}1    
}
\begin{document}

\title[ProjecTA: A Semi-Humanoid Robotic Teaching Assistant with In-Situ Projection for Guided Tours]{ProjecTA: A Semi-Humanoid Robotic Teaching Assistant with In-Situ Projection for Guided Tours}
\subtitle{To appear at ACM CHI ’26.}


\author{Hanqing Zhou}
\authornote{Both authors contributed equally to this research.}
\orcid{0009-0004-2077-6030}
\affiliation{%
  \institution{School of Design, SUSTech}
  \city{Shenzhen}
  \country{China}
}
\email{12331483@mail.sustech.edu.cn}

\author{Yichuan Zhang}
\authornotemark[1]
\orcid{0009-0004-1231-7686}
\affiliation{%
  \institution{School of Design, SUSTech}
  \city{Shenzhen}
  \country{China}
}
\email{12531639@mail.sustech.edu.cn}

\author{Zihan Zhang}
\orcid{0009-0001-1452-4388}
\affiliation{%
  \institution{School of Design, SUSTech}
  \city{Shenzhen}
  \country{China}
}
\email{zhangzihan654@gmail.com}

\author{Wei Zhang}
\orcid{0000-0002-1552-1365}
\affiliation{%
  \institution{School of Psychology, Shenzhen University}
  \city{Shenzhen}
  \country{China}
}
\email{zhangwei633@szu.edu.cn}

\author{Chao Wang}
\orcid{0000-0003-1913-2524}
\affiliation{%
  \institution{Honda Research Institute Europe}
  \city{Offenbach/Main}
  \country{Germany}}
\email{chao.wang@honda-ri.de}

\author{Pengcheng An}
\authornote{Corresponding Author.}
\orcid{0000-0002-7705-2031}
\affiliation{%
  \institution{School of Design, SUSTech}
  \city{Shenzhen}
  \country{China}
}
\email{anpc@sustech.edu.cn}

\renewcommand{\shortauthors}{Zhou et al.}

\begin{abstract}

  Robotic teaching assistants (TAs) often use body-mounted screens to deliver content. In nomadic, walk-and-talk learning, such as tours in makerspaces, these screens can distract learners from real-world objects, increasing extraneous cognitive load. HCI research lacks empirical comparisons of potential alternatives, such as robots with in-situ projection versus screen-based counterparts; little knowledge has been derived for designing such alternatives. We introduce ProjecTA, a semi-humanoid, gesture-capable TA that guides learners while projecting near-object overlays coordinated with speech and gestures. In a mixed-method study (N=24) in a university makerspace, ProjecTA significantly reduced extraneous load and outperformed its screen-based counterpart in perceived usability, usefulness of visual display, and cross-modal complementarity. Qualitative analyses revealed how ProjecTA’s coordinated projections, gestures and speech anchored explanations in place and time, enhancing understanding in ways a screen could not. We derive key design implications for future robotic TAs leveraging spatial projection to support mobile learning in physical environments.
  
\end{abstract}



\received{20 February 2007}
\received[revised]{12 March 2009}
\received[accepted]{5 June 2009}


\begin{teaserfigure}
  \includegraphics[width=\textwidth]{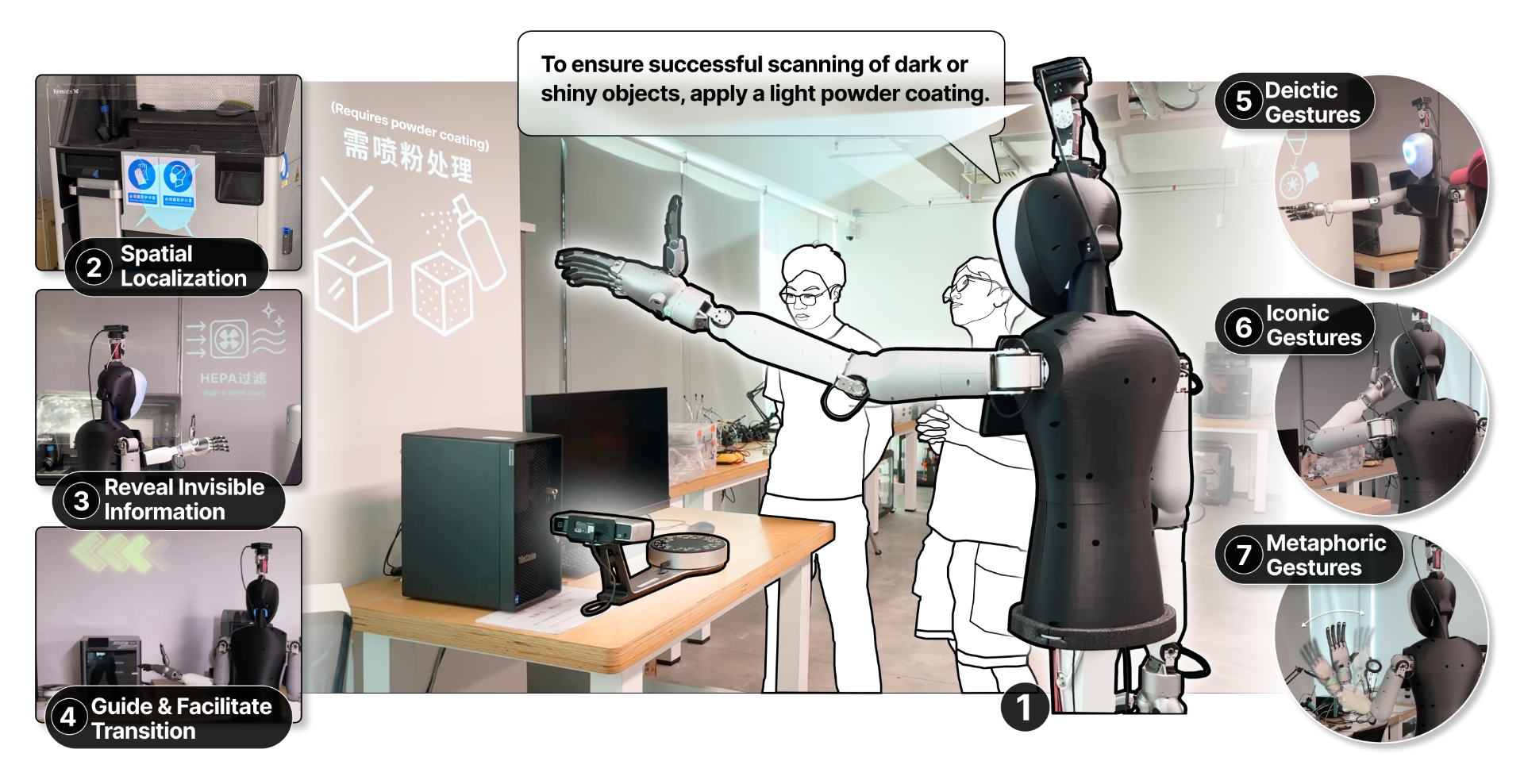}
  \caption{\protect\circled{1} System overview of ProjecTA, offering guided tours in a makerspace. Through in-situ projection, ProjecTA \protect\circled{2} provides clear spatial localization, \protect\circled{3} visualizes hidden equipment information, and \protect\circled{4} facilitates transitions. It employs gestures, including \protect\circled{5} deictic gestures pointing to concrete objects or locations, \protect\circled{6} iconic gestures depicting shapes or operations, and \protect\circled{7} metaphoric gestures conveying abstract ideas.}
  \Description{A composite image of the humanoid robot ProjecTA in a workshop. The main image shows the robot gesturing towards the ceiling while interacting with two people depicted as line-drawings. A speech bubble from the robot reads, "To ensure successful scanning of dark or shiny objects, apply a light powder coating." Seven numbered insets show close-ups of the robot's behaviors: projecting highlight halos to indicate safety symbols, projecting information and arrows on a wall, and performing various hand gestures like pointing, a hand-to-head covering pose, and a waving motion.}
  \label{fig:teaser}
\end{teaserfigure}

\maketitle

\section{Introduction}
Across formal and informal learning settings, robots serving as teaching assistants (TAs) can meaningfully offload repetitive and standardized tasks \cite{belpaeme2018social}. For instance, they can handle routine classroom logistics such as roll call, task reminders, and repeatedly deliver foundational explanations for novices \cite{bellas2024education}. They also provide step-by-step prompts \cite{yoshino2023teaching} and safety checks for hands-on procedures \cite{hasnine2024socially} in laboratory-like environments. And beyond classrooms and labs, they show potential in supplementing background facts or maintaining tour pace in exhibitions or museums \cite{ro2019projection,jiang2025visiobo}. 

Many deployed Robotic TAs have upper-torso or fully humanoid forms, ranging from small, desktop units (e.g., NAO \cite{baxter2017robot,ramachandran2019personalized}) to larger, mobile robots capable of gesturing (e.g., Pepper \cite{tanaka2015pepper,yoshino2023teaching}). 
A key motivation for upper-torso humanoid platforms is their ability to leverage human-like nonverbal signals, especially gestures, which support learner's comprehension and improve instructional task performance \cite{hostetter2011gestures,michaelis2022embodied}. 
For instance, pairing robot speech with co-speech gestures or on-screen cues improves word learning in child L2 tutoring \cite{demir2020l2}; and robot pointing helps learners locate targets faster and reduce misunderstandings \cite{sauppe2014robot}. 

To pair speech with visuals, current robotic TAs typically rely on chest- or head-mounted displays during face-to-face interaction. For example, robotic TAs utilizing the Pepper platform can present instructional content, such as supplementary visuals and structured learning points, on its chest-mounted tablet  \cite{yoshino2023teaching,tanaka2015pepper,sievers2025humanoid}. 
Such a screen-based paradigm is especially beneficial for close-range, face-to-face, and robot-centric engagement where the robot and its on-board screen are the learners' primary visual focus during the moment of interaction \cite{tanaka2015pepper,yoshino2023teaching,baxter2017robot,ramachandran2019personalized}.


However, robotic TAs are also needed in nomadic contexts, where learners and robots, instead of interacting face-to-face, move through space and jointly engage with external targets distributed across different spots. We refer to this arrangement as \textit{nomadic learning}, which is commonly seen in settings such as museums ~\cite{burgard1999experiences,gehle2017open,iio2020human}, instructional labs \cite{suddrey2018enabling}, or makerspaces.
Here, a robot's screen can distract learners' focus from the real-world object of interest, causing attention switching between the display and the physical referent. According to Cognitive Load Theory, such mental integration of separate information sources can increase extraneous cognitive load for learners \cite{ChandlerSweller1991}.

To support learners in such contexts, the robot, instead of being the center of attention, should be a coordinator, directing focus toward environmental targets and anchoring information around them \cite{rosen2019communicating}. Prior work shows that in-situ augmentative displays, enabled via headsets or stationary projectors, can reduce attention switching and lower cognitive load by co-locating information with its physical referent \cite{rosen2019communicating,ganesan2018better,bimber2005spatial}.
Recognizing this potential, a few design cases combined (non-humanoid) robotic chassis with on-board projection \cite{jiang2025visiobo,ro2019projection}, suggesting benefits of supplementing speech with projected content, while leaving the gesture-projection integration an unaddressed opportunity. 

To date, HCI research lacks empirical comparisons to examine whether robotic TAs with in-situ projection would outperform their screen-based counterparts in nomadic learning, how they may shape learners' experiences, or what design implications follow. 

Addressing these opportunities, we present \textit{ProjecTA}, a mobile robotic TA offering novice learners on-boarding tours in a makerspace. ProjecTA is equipped with a head-mounted projector, an arm-and-hand actuation module, and a mobile chassis (see \autoref{fig: hardware}). As shown in \autoref{fig:teaser}~\circled{1}, it places near-object overlays in coordination with voice narration and gestures. ProjecTA prepares its tours via a presentation-choreography workflow, which translates teaching goals into executable scripts orchestrating segmented speech, visual display assets, and callable gesture units on a unified timeline.
We evaluated ProjecTA via a mixed-methods, within-subject comparative study in a university makerspace, in which 24 participants completed guided tours using ProjecTA and its functionally equivalent counterpart with a chest-screen display, called \textit{Baseline}. Our research question (\textbf{RQ}) is: \textbf{\textit{How does a robotic TA with in-situ projection, compared with a screen-based counterpart, affect learners’ experiences during makerspace tours?}}

Rich empirical data yielded from our mixed-methods evaluation show that quantitatively, ProjecTA significantly lowered learners' extraneous load and resulted in higher perceived usability, usefulness of visual display, and cross-modal complementarity in comparison with Baseline. 
The qualitative findings further contextualize how the near-object visual overlays projected by ProjecTA reduced referent matching and attention switching in the space, offered a spatially accessible display to nomadic learning, and visualized equipment's critical or hidden details on-site. 
Concrete examples reveal how the robot's projection, gestures, and speech supplemented one another: for instance, projected visuals disambiguated pointing gestures, and iconic gestures reinforced visual and verbal messages.
Relevant design implications are thereby derived to inform future practice.

This work thereby contributes: (1) ProjecTA, a gesture-capable robotic TA with in-situ projection for supporting learners in guided tours; 
(2) an empirical comparison between such a system and a screen-based counterpart, revealing its positive impacts, and offering rich qualitative accounts of the learners' experiences;
(3) a set of design implications for future robotic TAs with in-situ projection to facilitate nomadic learning in physical settings.

\section{RELATED WORK}

\subsection{Robotic Teaching Assistants in Physical Learning Environments}
In HCI and educational technology, an increasing body of work has examined robots functioning as teaching assistants in real-world learning environments such as classrooms \cite{yoshino2023teaching,baxter2017robot,demir2020l2,bellas2024education} or supporting place attachment \cite{hu2025designing}. Prior studies have shown the feasibility of introducing robots in classroom environments, and proven benefits for both educators and learners \cite{rosanda2019robot}. Literature views robotic TAs as assistants rather than replacements: they offload repetitive, standardized routines (e.g., offering roll call, reminders, or repeated knowledge explanations for novices), hence allowing educators to focus on integrative and empathetic work \cite{belpaeme2018social,mubin2013review}.

Many deployed robotic TAs take a fully or upper-torso humanoid form, typically combined with an on-robot display pairing the robot speech with visuals, which is suited for close-range, face-to-face interactions. For example, Pepper pairs speech with a chest-mounted tablet to present instructional content and menu-style prompts \cite{tanaka2015pepper,yoshino2023teaching}; Furhat uses a head-mounted, back-projected face that supports gaze and lip-sync while placing short textual or graphic prompts near the head \cite{al2012furhat}. QTrobot is a small screen-face cartoon tabletop humanoid used as a peer to nudge children's handwriting posture \cite{wang2024co}. More broadly, desktop humanoids such as NAO are typically positioned within arm's length and synchronized with a nearby screen for stepwise tasks or quizzes, emphasizing near-field scaffolding \cite{baxter2017robot,ramachandran2019personalized}. These platforms are thus optimized for deskside tutoring and proximal face-to-face interactions.

However, this on-body screen design presents a key challenge: it forces learners to look at the robot instead of the physical object being discussed. Cognitive Load Theory (CLT) \cite{ChandlerSweller1991} predicts a split-attention effect when verbal explanations and related visuals are spatially separated, increasing extraneous load; co-locating cues on or next to the referent mitigates this cost \cite{ChandlerSweller1991}.
While empirically effective for deskside Q\&A \cite{rosenberg2020robot,rosenberg2019human}, the on-body screen solution can degrade during multi-object walkthroughs, where learners must shift their gaze between the artifact and the on-body screen \cite{lozada2025anywhere}. This motivates moving visuals off the robot and onto the target in object-centered, nomadic teaching.

We use `walk-and-talk' to describe instruction delivered while moving through a space and stopping at relevant items. In museums, early guides such as RHINO navigated galleries and stopped at exhibits for explanations, guiding visitors re-orienting to new artifacts along the route \cite{burgard1999experiences}. A more recent museum robot proactively approaches visitors for in-situ explanations \cite{iio2020human,gehle2017open}. In instructional labs, mobile humanoids like Pepper traverse benches and stations, with the group re-forming around different pieces of equipment \cite{suddrey2018enabling}. These learning settings are thereby object-centered and spatially distributed.

In such settings, a robot is more effective as a mobile coordinator than a screen: it should direct attention and provide information directly in the environment. 
This can be achieved in two ways: first, by using deictic gestures such as pointing, to establish joint attention on a specific object \cite{sauppe2014robot}; second, by presenting information as in-situ visuals, such as explanatory overlays projected onto the object itself.
This approach can mitigate the split-attention effect, aligning with both CLT and HRI findings on the benefits of in-place visualization \cite{ChandlerSweller1991,rosen2019communicating,ganesan2018better}. However, most empirical evidence comes from stationary operations such as assembly and repair tasks in fixed locations, where evaluation mainly focused on task performance \cite{tang2003comparative,aschenbrenner2019comparing,rosenthal2010augmenting}, rather than on learners’ extraneous cognitive load. A few in-vehicle navigation studies assessed attentional distraction and showed that head-up displays (HUDs) can reduce glances and divided attention compared with a conventional dashboard screen \cite{kim2009simulated}, but these settings still involve seated drivers rather than learners moving through space. Thus, HCI still lacks empirical evidence on whether and how in-situ projection, in comparison with conventional screen-based displays, can influence learners' extraneous cognitive load in nomadic learning with mobile robots.

As instruction shifts from deskside tutoring to walk-and-talk tours, challenges of screen-bound humanoids become clear: first, physical gestures are often not precise enough to identify distant or out-of-view objects, leading to confusion about what is being referenced  \cite{sauppe2014robot,huang2024gestures}. 
Pairing speech with on-body display separates auditory and visual information from the target object, increasing extraneous cognitive load for learners \cite{ChandlerSweller1991}. Third, body-mounted screens are difficult for learners to see from different angles while moving, hindering shared learning experience \cite{lozada2025anywhere}.
These constraints motivate moving visuals off the robot and into the shared environment via public, in-situ augmentation.

\subsection{Public Augmentative Displays in Learning Settings}
Public augmentative displays present information in the environment, so co-located learners can see and reference it together \cite{an2020ta}, using it as an anchor for discussion, coordination, and sense-making \cite{radu2014augmented,barthele2023exploring}. 
Pedagogically, this approach supports joint attention and efficient reference resolution \cite{clark1991grounding,tomasello1986joint}; and with information co-located with its referent, learners avoid mentally integrating separated sources, reducing extraneous cognitive load \cite{ChandlerSweller1991}.

In classrooms, such shared interfaces typically appear as wall displays \cite{soneral2017scale,van2015lernanto,gordy2018technological}, fixed projection \cite{caldwell2007clickers,moraveji2011classsearch}, and interactive whiteboards \cite{mercer2010using,hennessy2011role}: for instance, visually encoding students' progress on a wall display to build shared awareness \cite{van2015lernanto}, and projecting learners' ongoing web searches to facilitate timely feedback and whole-class discussion \cite{moraveji2011classsearch}. Beyond classrooms, museum exhibits like DeepTree use large, shared surfaces for collaborative exploration \cite{block2012deeptree}; and an industrial training system projects step-by-step instructions onto a workpiece to guide novices \cite{buttner2020augmented}.
Across these settings, such public augmentative displays speed learners' shared access to key information and help them coordinate.

Since public augmentative displays are shared by default, any nearby learner can see, point to, and reference without extra individual equipment (e.g., wearables or handhelds). 
For instance, AAR system uses an actuated projector to help bystanders see the same AR content as HMD users \cite{hartmann2020aar}. Radu's meta-review on educational AR argues for the scaffolding effects of shared, in-room overlays in co-located learning settings \cite{radu2014augmented}.
By contrast, wearable or handheld AR systems provides individualized, private overlays by default, requiring mirroring or streaming when needed for public access \cite{lukowicz2015glass}. 
Such private displays benefit personalized guidance or asymmetrical collaboration \cite{gil2014ar,turakhia2025investigating}, such as Lumilo glasses which inform teachers for tailoring instructions to students' real-time needs \cite{holstein2018student}.
However, for maintaining learners' pace and delivering shared content in co-located settings, co-attendable public displays are more convenient, without requiring learners wearing and holding extra devices \cite{hartmann2020aar}.

However, most public displays are fixed installations (wall-mounted screens, ceiling projectors, etc.). Their utility can be limited in nomadic and spatially distributed learning. In settings like a museum tour or a lab walkthrough, where the focus of instruction moves from one object to another, a fixed display cannot follow the learners.
This motivates our exploration of a nomadic, in-situ approach and how it can better support object-centered, walk-and-talk learning.

\subsection{Combining In-situ Projection with Robotic Systems}

In-situ projection is proven to support fixed-location tasks via information overlays onto the area of operation.
For instance, ImproVisAR overlays a `piano roll' onto a keyboard as intuitive guidance \cite{deja2025improvisar}, while LuminAR turns a desk into an interactive surface \cite{linder2010luminar}.
Evidences from assembly-style tasks show that in-situ projection lowers workload and improves performance relative to paper/tablet/HMD baselines (Pick-by-Projection \cite{baechler2016comparative}; LEGO setups \cite{funk2016interactive}). 
While most of these evidences come from fixed workstations or tabletops, empirical comparisons in educational contexts, especially nomadic, walk-and-talk learning, remain scarce. 
Leppink’s Cognitive Load Scale (CLS) \cite{leppink2013development} distinguishes learners' Extraneous load (avoidable effort due to presentation) from Intrinsic Load (inherent complexity of learning topics). Since in-situ projection co-locates cues with their referents, it mitigates split-attention: a primary source of extraneous load \cite{chandler1992split}. 
Our study thereby uses CLS \cite{leppink2013development} to test whether in-situ projection, integrated in nomadic settings, would lower learners' extraneous load, extending benefits reported at fixed workstations \cite{baechler2016comparative,funk2016interactive}.

In robotics, external, fixed projectors have been used to augment robots and their workspace. For instance, ShapeBots uses a ceiling-projector over a tabletop micro-robot swarm for data physicalization \cite{suzuki2019shapebots}.
RobotIST overlays IDE-style feedback (errors/next step/state) on the task surface beside a desktop robotic manipulator, easing procedures more than conventional tools \cite{sefidgar2018robotist}.

Beyond fixed setups, on-board projectors have been mounted to actuated gimbals, mobile chassis, or drones to make the display angle-adjustable or fully mobile \cite{suzuki2022augmented}.
PAMI, a stationary meeting-room installation, orients its projected overlays to nearby surfaces for hybrid collaboration \cite{ro2018pami}. 
A few systems attach projectors to mobile chassis to deliver near-object overlays in museums \cite{ro2019projection} and poster exhibitions \cite{jiang2025visiobo}. 
Another system navigates indoors and projects an on-demand interactive surface on a chosen spot around exhibits \cite{elsharkawy2021uwb}. 
In children's play settings, a wheeled pro-cam robot roams the room and projects scene backdrops and prompt cues on the floor/props to guide dramatized activities \cite{ahn2011projector}. 
Many mobile robots also project arrows, paths, or footprints onto nearby surfaces to convey motion intent and planned trajectory
\cite{watanabe2015communicating,coovert2014spatial,chadalavada2015s,han2022projecting}. 
Related work on mobile projection mapping keeps overlays geometrically aligned while moving \cite{kasetani2015projection}, and drone-based projection enables pop-up displays where installation is impractical \cite{darbar2019dronesar,lingamaneni2017dronecast,scheible2013displaydrone}.
While these systems effectively make visuals mobile and situated, projection is typically utilized as a single output channel.
One exception, Visiobo, times projected visual overlays with LLM-based auditory narration \cite{jiang2025visiobo}, showing the promise of multimodal coordination. Yet prior systems mostly confine co-reference to speech and visuals, with almost no exploration of jointly coordinating gestures, in-situ projection, and narration. In particular, the integration of in-situ projection with gesture-capable robots, such as upper-torso or humanoid platforms, remains unaddressed.

Apart from projection-specific systems, a large body of work have studied multimodal coordination in social robotics.
GenComUI synchronizes generated visual aids (map annotations, route cues, animated feedback) with spoken instructions, outperforming speech-only baselines \cite{ge2025gencomui}. ELEGNT pairs expressive robot motion with directional lighting/visual signals, making intent more legible and boosting trust \cite{hu2025elegnt}. 
Zhang et al. use gaze and environmental context to ground indirect requests, improving team coordination \cite{zhang2025can}. 
Leusmann et al. coordinate questioning with exploratory actions to sustain engagement \cite{leusmann2025investigating}. 
Most prior work assumes close-range, face-to-face interaction with information centered on the robot, leaving open how to design the communication channels when the robot and people jointly engage an external object in non-face-to-face, walk-and-talk learning. 

Building on the potential benefits of in-situ projection \cite{funk2016interactive,chandler1992split}, while retaining the instructional value of gestures \cite{clark2004speaking,cook2008gesturing}, we explore a gesture-capable robot with an on-board projector that coordinates in-situ projection, arm-and-hand gestures, and speech to support learners in nomadic, walk-and-talk physical learning environments.

\begin{figure*}[tb]
  \centering
  \includegraphics[width=\textwidth]{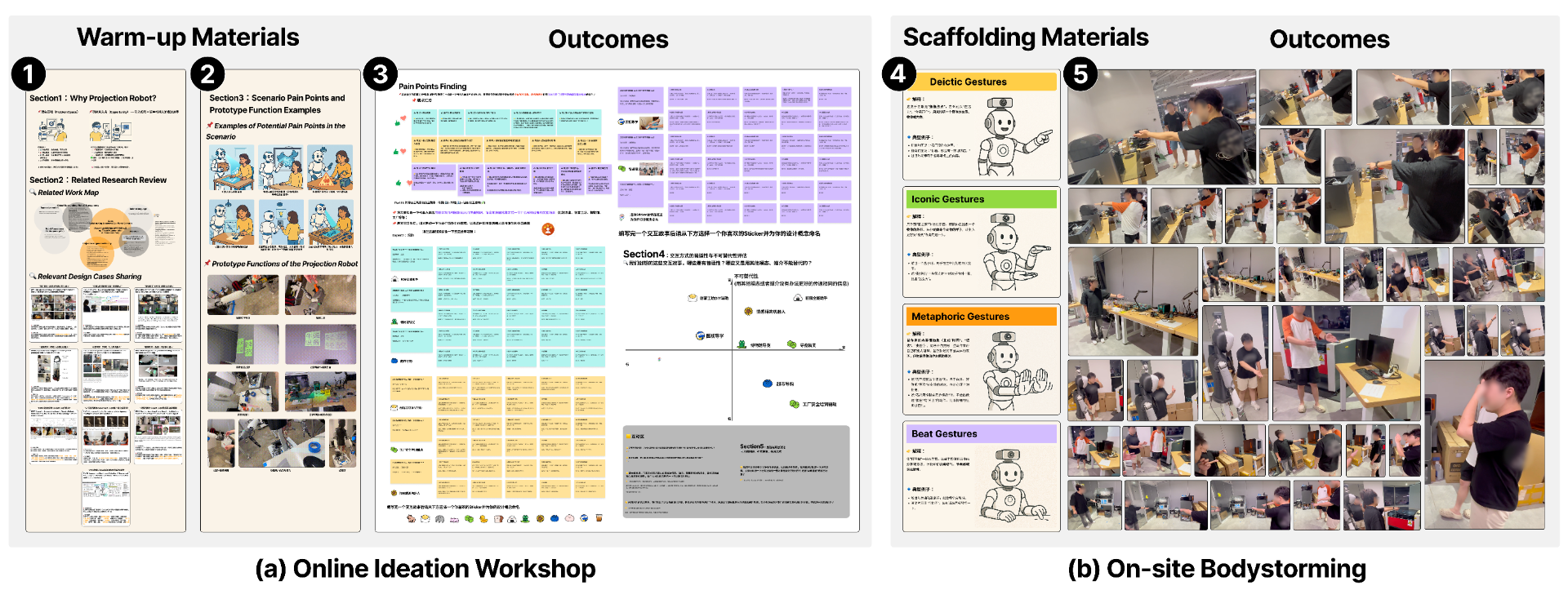}
  \caption{Formative studies: (a) warm-up materials and outcomes from the online ideation workshop; (b) scaffolding materials and outcomes from the on-site bodystorming.}
  \Description{A two-part figure illustrating the materials and outcomes of formative studies. Panel (a) shows materials and outcomed from an online ideation workshop, including presentation slides with diagrams, text, and cartoons. Panel (b) displays materials and outcomes for on-site bodystorming, such as illustrated gesture definitions, alongside a grid of photos showing participants performing these gestures in a real makerspace.}
  \label{fig: formative_study}
\end{figure*}

\section{Formative Study}
To explore the value and design opportunities of pairing a teaching assistant (TA) robot with in-situ projection to facilitate physical learning environments, we conducted two co-design activities in our formative study: (1) an online ideation workshop with experts in Robotics and HCI to broadly identify design opportunities; and (2) a bodystorming session in a makerspace to ground our design and collect contextualized requirements.

\subsection{Online Ideation Workshop with Robotics and HCI Experts}

\subsubsection{\textbf{Participants and Procedure}}
We recruited six experts (E1–E6). E1 was the Chief Scientist at a Robotics research institute in Europe, and E4 was a university professor in Robotics in Asia, while the remaining experts held master's degrees in Robotics~(E3, E5, E6) or HCI~(E2). Participants were split into two groups. Each group used Figma \footnote{\url{https://www.figma.com/}} to collaborate while having an online meeting facilitated by the researchers, lasting around 2.5 hours.

Each group was offered a Figma template aligned with the co-design agenda to scaffold their collaboration (see \autoref{fig: formative_study}~(a)). At the start, the experts were briefed about the core objective of the session, i.e., to identify design opportunities for robotic TAs with in-situ projection. As warm-up materials (see \autoref{fig: formative_study}~\circled{1}), relevant design cases from HCI and Robotics on robots presenting information in physical environments were briefly summarized for the experts. These cases were meant to help them understand the background and spark innovations that both build upon and diverge from previous solutions. To ground the discussion, we also shared short demo videos showcasing our robot so experts could understand the functional capabilities of the hardware platform (see \autoref{fig: formative_study}~\circled{2}). 

In the next phase, each expert was asked to broadly ideate As-Is scenarios where robots with in-situ projection might be helpful~(20 min). Building on these As-Is scenarios, they then created solution-oriented To-Be scenarios~(30 min). To facilitate experts' concrete To-Be solutions~\cite{koike2024tangible}, a structured format prompted entries on how the robot should coordinate in-situ projection, gesture, and voice, and how it should interact with actors and objects in the space.

In a follow-up discussion (30 min), the experts were asked to heuristically evaluate all the created ideas on two aspects: the unique advantages of in-situ projection within each scenario and how these advantages might generalize to other scenarios.
Finally, the workshop concluded with a 10-minute collective review, the co-design outcomes were shown in \autoref{fig: formative_study}~\circled{3}.

\subsubsection{\textbf{Major Insights}}
The experts contributed thirteen detailed To-Be scenarios with rich, vivid commentaries. We analyzed these outcomes using an affinity diagram \cite{lucero2015using}.
The resulting insights, serving as our early design inputs, helped us identify a set of key Design Opportunities (\textbf{DO1-DO4} below) that in-situ projection may open up for robotic TAs, as well as typical design scenarios to begin with. We summarize the major insights below: 

\textbf{DO1- Pinpointing key objects and locations through spatial cues}: In several To-Be scenarios for in-situ learning or operational tasks, the experts noted that robots' pointing sometimes may be coarse or vague when needing to refer to regions or objects through the space. They envisaged using in-situ projection to anchor clearer references in the scene. 
\textbf{Examples:} E5 envisioned a projection robot highlighting key regions of posters or exhibits, compensating a physical pointing. E3 envisaged a robotic TA overlaying projected positional cues on a workpiece to indicate where to apply the power drill, in a furniture-repairing scenario.

\textbf{DO2- Presenting learning content alongside its physical referent}: The To-Be scenarios related to training or touring highlighted that in-situ projection could place learning materials beside the objects being described simultaneously by a robotic TA. This could avoid learners constantly looking back and forth between an object and its explanation. \textbf{Examples:} E5 envisaged that in nursing training, a robotic TA could project task briefings, key steps, etc., next to the mannequin, allowing trainees to glance-and-act. Similarly, E6 envisaged a robotic TA projecting co-located explanations for learners in a makerspace.

\textbf{DO3- Projecting visuals to complement or summarize robotic TAs' speech}:
Experts' To-Be scenarios also envisioned that in-situ projection could visualize concise summaries or add-on materials that complement robots' verbal outputs on the spot. \textbf{Examples:} E4 envisaged a museum tour, where a robot projected background information and auxiliary materials at places along the way, deepening visitors' understanding. E2 described a similar example: a robot using projection to consolidate its verbalization in a poster exhibition.

\textbf{DO4- Visualizing hidden or invisible information on-site}: 
A few To-Be scenarios imagined how in-situ projection could help robotic TAs make hidden or occluded content visible to users. \textbf{Examples:} In E1's example of organizing items in a domestic setting, a robot projected information onto a nearby wall to remind the user of misplaced items and items stored out of sight, which would otherwise remain invisible to the user.

The To-Be scenarios created by the experts consistently highlighted the most distinctive potential of a robotic TA with in-situ projection: it affords unobtrusive, ad-hoc, on-site information aid for users to both learn about and learn with spatial and physical objects in a particular environment (e.g., museum, makerspace, workshop, exhibition...). We chose one of such typical scenarios: a robotic TA guiding beginners through an introductory tour of a makerspace, which the experts identified as a particularly rich and representative setting even though museum-like environments (e.g., museums\cite{ro2019projection} and exhibitions\cite{tamai2019method,jiang2025visiobo}) have received more prior attention in related work. In line with the experts’ feedback, a makerspace contains various pieces of equipment distributed spatially, each requiring specific instructional materials, suggesting sufficient space to explore all four design opportunities summarized above. For onboarding in a makerspace, each beginner often needs a similar introductory tour, echoing a typical duty of robotic TAs: offloading repetitive, standardized tasks from educators. For these reasons, the beginner onboarding tour in a makerspace has been chosen as our design scenario to proceed with.

\subsection{On-site Bodystorming with Makerspace Experts}
To collect contextualized requirements and practical inputs for designing a robotic TAs' speech, gestures, and in-situ projection, we conducted another co-design session in a university makerspace(see \autoref{fig: formative_study}~(b)), adopting a bodystorming method, which has been proven to offer on-the-ground design ideas with situational factors by having participants enact and role-play in the context \cite{oulasvirta2003understanding}. 

\subsubsection{\textbf{Participants and Procedure}}
We recruited a makerspace educator and manager (M1) and two experienced makers (M2, M3), each with more than 6 years of active engagement in makerspace environments. Under researcher facilitation, they completed a 35-minute in-situ bodystorming session, where they took turns to role-play a robotic TA offering novices an introductory tour, acting out its speech, gestures, and visual presentation using both bodily performance and verbal description. Such enactment was meant to extract the experts' embodied know-how of how to introduce the makerspace, and gather references, examples, and inspirations for creating the robot's behaviors in different communicative channels.

Scaffolding materials were prepared to further support experts' enactment. First, to help experts comprehensively enact potential gestures, we provided them with a set of gesture classification cards (see \autoref{fig: formative_study}~\circled{4}) based on McNeill's classification~\cite{mcneill1992hand} as widely used in sociology and educational sciences to analyze interpersonal communication (e.g., in classroom teaching or social conversing).
The cards provide clear definitions and examples of each gesture type: (1) Deictic (pointing to entities or locations), (2) Iconic (depicting concrete objects or actions; e.g., holding both hands apart to indicate a part's width), (3) Metaphoric (mapping abstract concepts onto space; e.g., an upward hand sweep to indicate "increase"), and (4) Beat (small rhythmic strokes aligned with speech). 

\begin{figure*}[tb]
  \centering
  \includegraphics[width=0.6\textwidth]{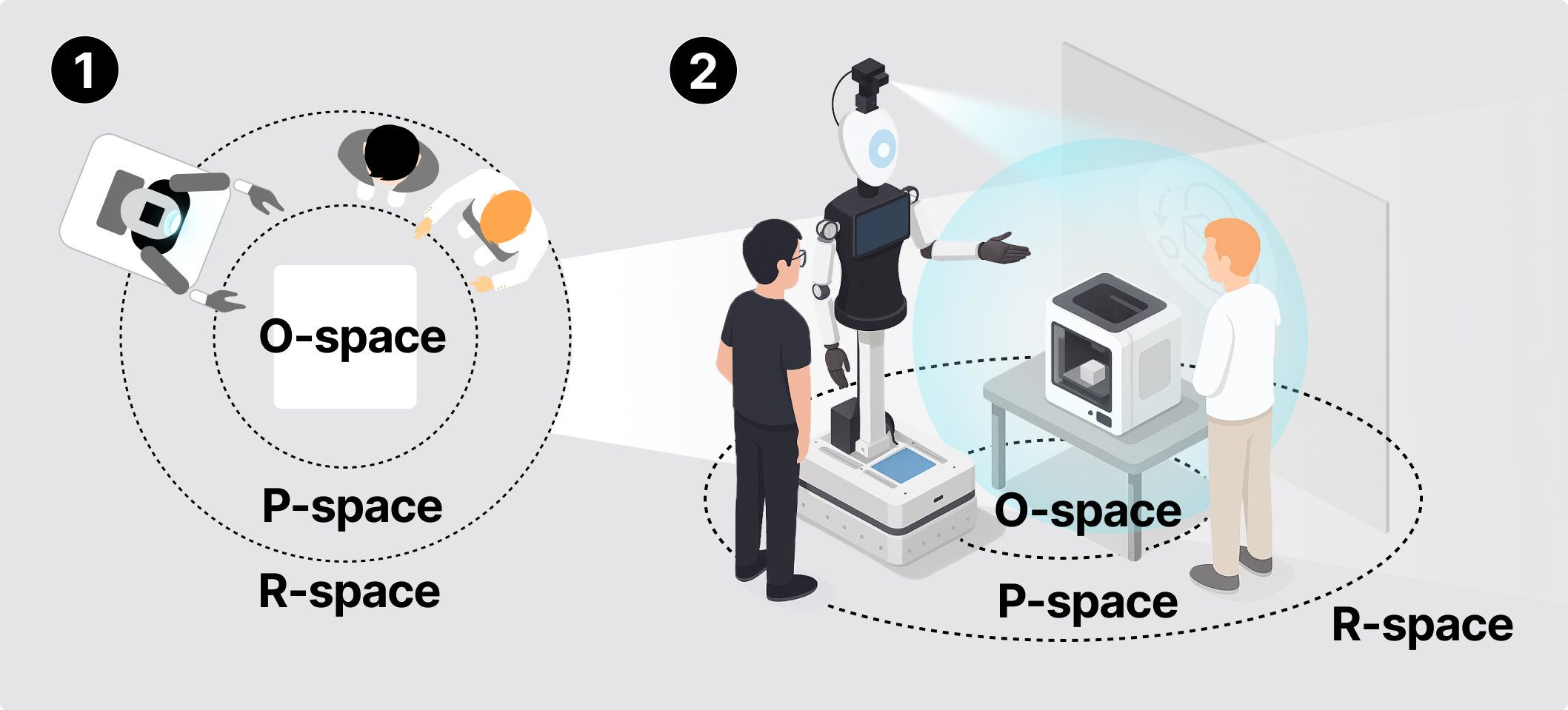}
  \caption{\protect\circled{1} Illustration of F-formation; \protect\circled{2} F-formation applied to in-situ projection placement in the makerspace.}
  \Description{A two-part figure illustrating the F-formation theory and its application in human-robot interaction. The first part, labeled 1, is a top-down diagram showing three concentric areas around a shared focus: an inner O-space, a middle P-space, and an outer R-space. The second part, labeled 2, shows this theory applied: a humanoid robot and two people interact around a 3D printer, with the three spaces mapped onto the scene as the robot projects information onto a surface.
}
  \label{fig: Fformation}
\end{figure*}

To help experts enact the positional relationships between the robot and the learners, we drew on the theory of F-formation~\cite{marquardt2012cross}, which conceptualizes spatial arrangements in interpersonal interaction (see \autoref{fig: Fformation}~\circled{1}) into three types of spaces: (1) O-space, a shared focus area in the center, where the device or object of interest should be located; (2) R-space, the surrounding interaction circle formed by the robotic TA and learners; (3) P-space, outsiders or bystanders remaining in the outer public space (not considered in our scenario). Related work has similarly leveraged F-formation to improve naturalness for robot–human engagement (e.g., the work by Yousuf et al. \cite{yousuf2013move}). In addition, we offered the experts an on-site live demonstration of the robot's prerecorded gestures and sample visual presentations to help them understand the hardware capabilities. 

The experts' enactment involved six makerspace devices (see \autoref{fig: formative_study}~\circled{5}). They visited each of the devices sequentially and enacted in turn the explanations they thought a robot should give to beginners. 
In each episode, they planned the key learning points, selected gestures (from the four types introduced earlier), described the visual presentation and its placement, and demonstrated how these outputs could coordinate with each other. 
Each enactment episode was followed by a brief discussion and on-the-spot iterations~\cite{segura2016bodystorming}.
The whole session was video-recorded for analysis.


\subsubsection{\textbf{Major Insights}}
The bodystorming session resulted in rich and concrete design inputs. For instance, the video clips of experts' bodily enactments were categorized based on McNeill's framework\cite{mcneill1992hand} for later constructing the robot's gestural unit library. The experts' verbal descriptions were subjected to a content analysis\cite{ford2004content} to extract key learning points of each device, as well as examples on how the robot's speech and visual display could be designed. To gain structured insights, we summarize all these inputs into six major design requirements (\textbf{DR1-DR6}):

\textbf{DR1- Verbal explanation should cover how it works, how to operate it, and safety information}: the experts' verbalization consistently indicated three types of learning points: principles of how the equipment works (e.g., forming principles of FDM and resin printers), basic methods of how to operate the equipment, and safety reminders (e.g., do not touch the soldering iron tip by hand). 

\textbf{DR2- Verbal explanation should show connections between devices and use clear transitions to move between them}: the experts' verbal articulation showed the need for coherent narration that integrates both content connections (such as comparing resin-based and FDM 3D printers and contrasting their forming principles) and clear transitions (e.g., \textit{``This equipment has now been introduced; please follow me to the next one''-M1}).

\textbf{DR3- Leveraging deictic gestures for orienting learners' attention in the makerspace}: the experts' bodily demonstrations showed how deictic gestures are essential for orienting learners' attention toward the intended equipment, area, or direction (as shown in \autoref{fig: formative_study}~\circled{5}), during a makerspace tour. Deictic gestures were also used to emphasize visual display content. For instance, when introducing the laser cutter, M1 said, \textit{``The operating procedure can be found in the image,''} while pointing toward the projection. Similarly, M3 suggested pointing to the visual display to make learners notice a learning point. In total, the experts' enactments generated 18 deictic gesture exemplars.

\textbf{DR4- Utilizing iconic and metaphoric gestures to vividly reinforce learning concepts}: the experts employed iconic gestures to mimic or illustrate specific objects or operations, thereby making verbal explanations more vivid. For example, using hand spanning or spacing to indicate workpiece size (M3) or 3D scanning distance (M2). Or covering eyes to warn against laser exposure (M1).
In addition, metaphoric gestures were enacted to convey and reinforce abstract concepts, such as waving to stress \textit{``don't leave waste''-M1}, or pushing hands forward to indicate prohibition (M2), or finger tapping to highlight cautions (M3). In total, experts generated 15 iconic and 12 metaphoric gesture exemplars on site, which we used to build the robot's gestural unit library.

\textbf{DR5- Visual displays should use simplistic and clear graphics and concise text to highlight key information}: the experts emphasized that the robot's visuals should remain simplistic and clear, avoiding lengthy text or overly complex illustrations. M3 envisaged that, unlike museum or exhibition tours, makerspace tours could benefit from visuals that resemble product manuals—using simple yet clear graphic styles. M2 stressed the importance of highlighting key information points, such as dimensions, distances, or temperatures, by directly labeling numbers on schematic diagrams. M1 further noted the value of visually emphasizing critical parts of the device, such as warning signs or the location of an emergency stop button.

\textbf{DR6- Visual displays need to be temporally aligned with the robot's speech and gestures}: through embodied enactments, the experts commonly emphasized that displayed visuals should appear in synchrony with the relevant narration and gestures, much like slides in a lecture. As M2 noted, when introducing operational steps, the robot should verbally describe each step while simultaneously presenting the corresponding image. M2 further suggested that sometimes, visuals could also follow the pointing gestures, such as highlighting a region with arrows while the hand gesture indicates the same area.

\section{The ProjecTA System}
In this section, we describe how the inputs from our formative study were translated into the design and implementation of the ProjecTA system.

\begin{figure*}[tb]
  \centering
  \includegraphics[width=\textwidth]{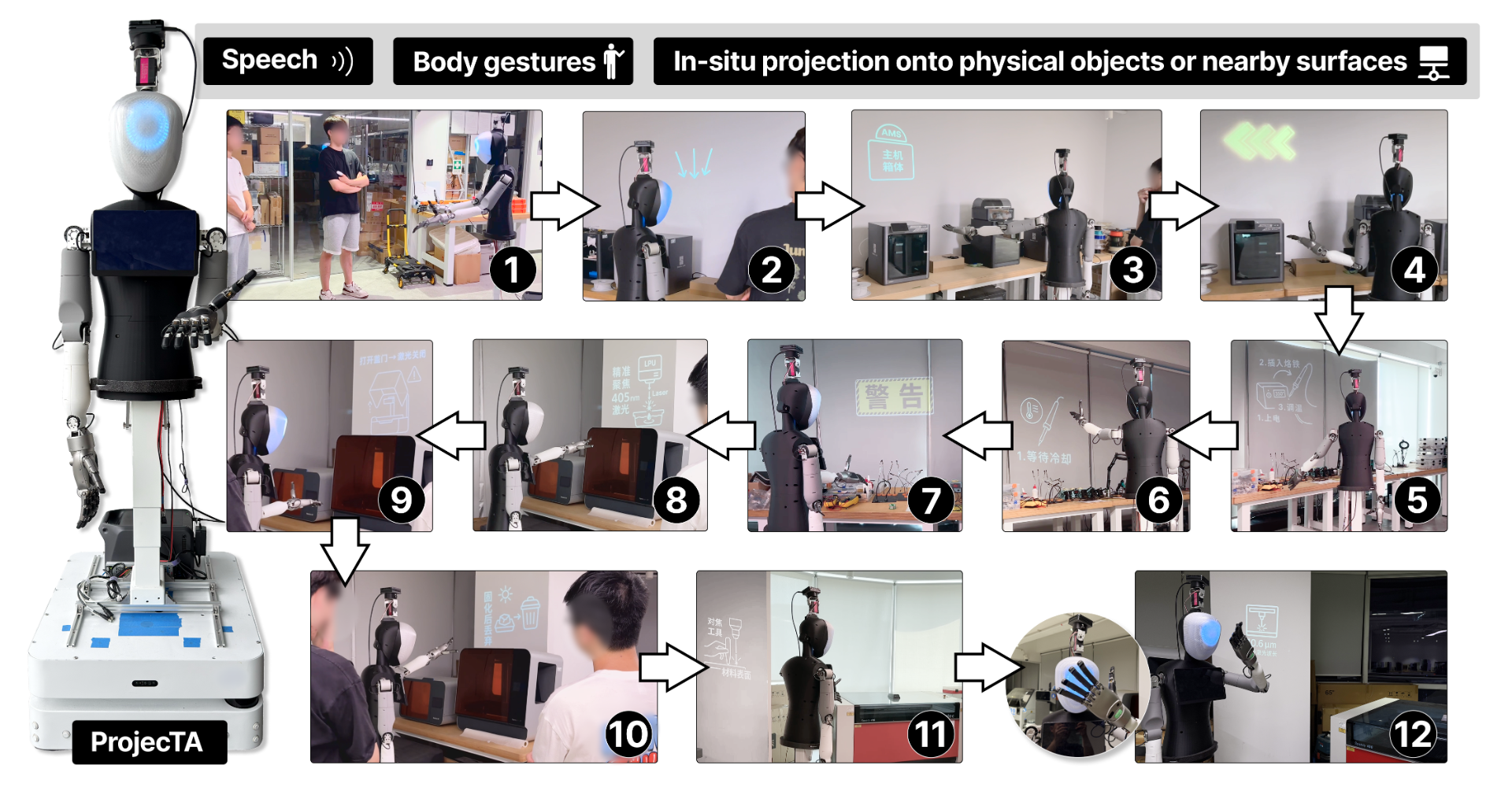}
  \caption{An exemplar usage scenario of ProjecTA during makerspace guided tours}
  \Description{A sequential diagram illustrating a 12-step guided tour led by the ProjecTA robot. The figure is laid out as a flowchart with twelve numbered panels that show a continuous interaction with a user in a makerspace. The sequence depicts the robot using its three key modalities listed at the top—Speech, Body gestures, and In-situ projection—to guide the person through tasks at various pieces of equipments. Examples shown include operating a 3D printer, using a soldering iron, and working with a laser cutter. In the panels, the robot is shown both physically pointing and projecting visual aids, such as instructional diagrams, arrows, and text in Chinese characters, directly onto near-by surfaces and equipment.}
  \label{fig: projecta_design}
\end{figure*}

\subsection{Design of ProjecTA}
ProjecTA is a prototype system meant to probe how a robotic TA with in-situ projection would support learners in a physical learning environment by offloading certain repetitive, standardized tasks from educational routines. In this study, ProjecTA has been specifically designed to offer novice learners guided tours in a makerspace. ProjecTA could present information through speech, body gestures, as well as in-situ projection onto physical objects or nearby surfaces in the space. 
As shown in \autoref{fig: projecta_design}, we illustrate its design details through an exemplar usage scenario:

Eager to explore the university makerspace, beginners Jeff and Jerry started a tour led by ProjecTA, a robotic guide.
At the entrance, ProjecTA greeted them (see \autoref{fig: projecta_design}~\circled{1}), giving them an overview of the tour before leading them to the northeastern corner, where several machines are placed in line.

To make the instructional target clear, ProjecTA projected an arrow above the FDM 3D printer\footnote{Fused Deposition Modeling: a 3D printer melts plastic filament and deposits it layer by layer to form an object.} (\textbf{DO1}, \autoref{fig: projecta_design}~\circled{2}). ``Let's look at the FDM 3D printer [...] Notice its two main components: the automatic material switching system on top and the chassis below.'' As the explanation unfolded, its projection displayed a simplistic and clear diagram (\textbf{DR5}, \autoref{fig: projecta_design}~\circled{3}): a semicircle and a square, labeled "Automatic Material Switching (AMS) System" and "Chassis," respectively. This helped the students quickly grasp the printer's structure.
Upon finishing, ProjecTA projected an animation of moving leftward to signal a transition, ``Next, we will look at the soldering station, please follow me.'' (\textbf{DR2}, \autoref{fig: projecta_design}~\circled{4})

At the soldering station, ProjecTA projected the operating steps right next to the soldering iron, so Jeff and Jerry could follow the instructions without looking away (\textbf{DO2}, \autoref{fig: projecta_design}~\circled{5}). When the robot described how to replace the soldering tip, the projected steps updated in real time. First, "Step 1: Wait for cooling," and then "Step 2: Loosen the nut," keeping the visual instructions synchronized with the verbal explanation (\textbf{DR6}, \autoref{fig: projecta_design}~\circled{6}). Afterward, the in-situ visuals switched to a warning symbol, ``[...] Beyond the basic function and operation,'' ProjecTA added, ``there are also safety precautions to keep in mind [...].'' (\textbf{DR1}, \autoref{fig: projecta_design}~\circled{7})

A directional projection then guided them to the next machine: the resin-based 3D printer.
``Unlike FDM printers, this machine cures resin layer by layer with a laser focused on the tank bottom,'' ProjecTA explained. 
As the curing process is a complex concept and cannot be demonstrated on the spot, the robot projected a diagram to complement and summarize its verbal explanation (\textbf{DO3}, \autoref{fig: projecta_design}~\circled{8}). 
Moving on, ProjecTA used pointing to direct the learners' focus to the printer's lid, explaining that opening this lid would automatically shut off the laser (\textbf{DR3}, \autoref{fig: projecta_design}~\circled{9}). Subsequently, the robot pointed to the newly projected visuals to stress the proper disposal of waste resin (\textbf{DR3}, \autoref{fig: projecta_design}~\circled{10}).

Finally, they stopped at the laser cutter, whose key component, the focusing lens, was not visible to visitors---it was too small and hidden inside the machine. Hence, ProjecTA projected a magnified illustration to show what this component looked like and how it worked (\textbf{DO4}, \autoref{fig: projecta_design}~\circled{11}). The robot continued its explanation. When it came to safety, the robot raised its hand in a shielding gesture in front of its eyes as a warning and cautioned, ``Don't look directly at the laser.''(\textbf{DR4}, \autoref{fig: projecta_design}~\circled{12})

The tour concluded with ProjecTA stating, ``This brings the session to an end. I hope the explanations have been helpful.'' Jeff and Jerry had completed their first learning experience in the makerspace.

\subsection{System Implementation of ProjecTA}

\begin{figure*}[tb]
  \centering
  \includegraphics[width=\textwidth]{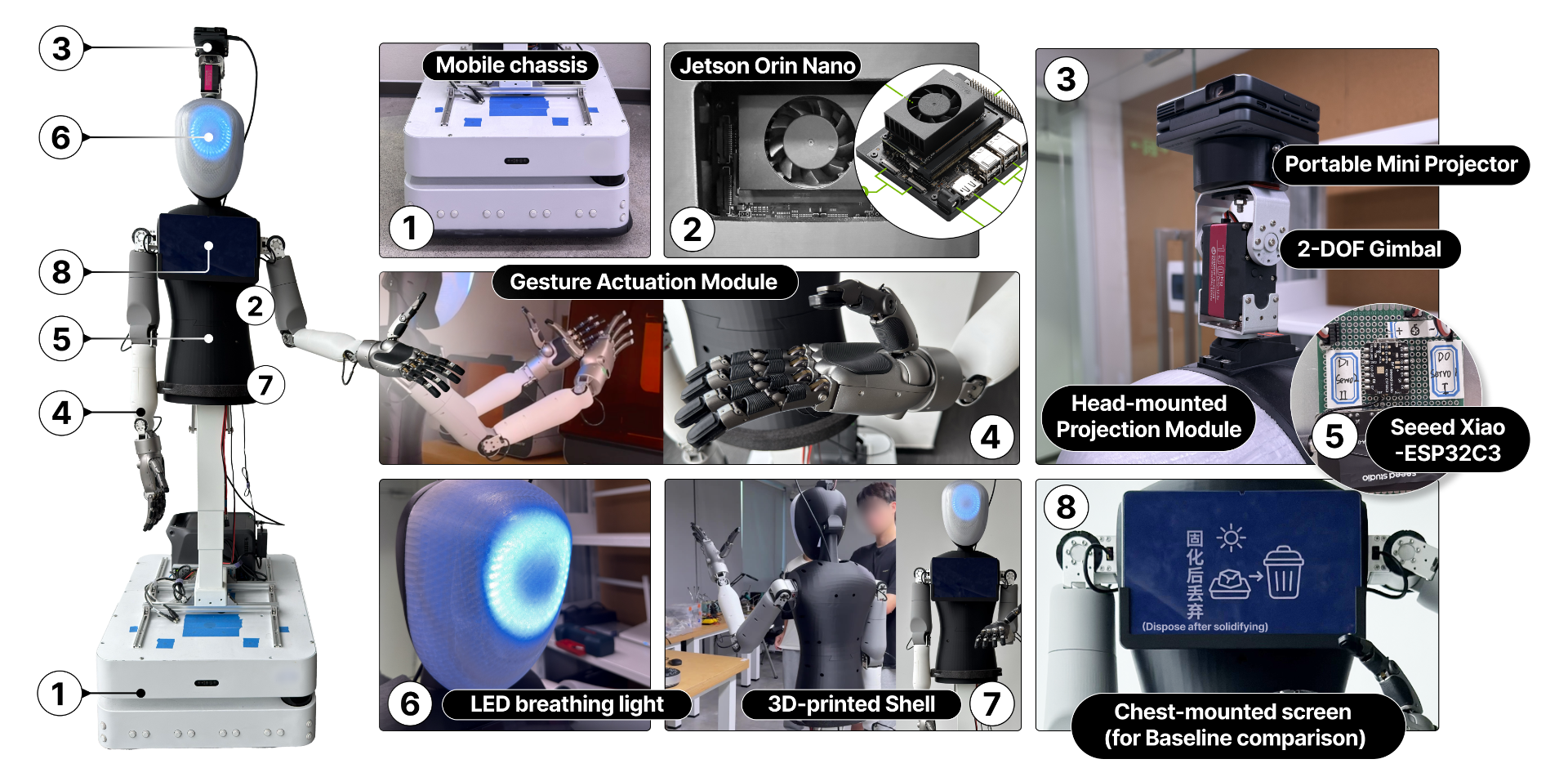}
  \caption{Hardware components of ProjecTA consist of a mobile chassis, a gesture actuation module, a head-mounted projection module, and a chest-mounted screen (for baseline comparison).}
  \label{fig: hardware}
  \Description{A diagram of the ProjecTA robot's hardware components, presented as a large central image of the robot with eight numbered callouts. Surrounding the main image are eight corresponding inset panels, each showing a labeled, close-up view of the part indicated by the callout number. The detailed components are: (1) the mobile chassis, (2) a Jetson Orin Nano computer board, (3) the head-mounted projection module, (4) the gesture actuation module showing the robotic arms and hands, (5) a close-up of the Seeed Xiao microcontroller, (6) the blue LED breathing light on the robot's head, (7) the 3D-printed outer shell, and (8) the chest-mounted screen.}
\end{figure*}

\subsubsection{\textbf{Robotic Hardware}}
As shown in \autoref{fig: hardware}, ProjecTA consists of a Gesture Actuation Module, a head-mounted projection module, and a mobile chassis. The Gesture Actuation Module integrates two 6-DOF\footnote{Degrees of freedom: the number of independent axes a mechanism can move or rotate around.} arms driven by Eyoubot\footnote{\url{http://www.eyoubot.com/}} planetary torque motors and a 20-DOF dexterous hand from Linkerbot\footnote{\url{https://linkerbot.cn/}}~(see \autoref{fig: hardware}~\circled{4}), controlled via a Jetson Orin Nano~(see \autoref{fig: hardware}~\circled{2}). 
The projection module carries an Aurzen ZIP Tri-Fold Portable Mini Projector\footnote{\url{https://aurzen.com/}} (720p, 100 ANSI lm, USB-C, auto-focus) mounted on a 2-DOF gimbal~(see \autoref{fig: hardware}~\circled{3}) powered by a Seeed Xiao-ESP32C3 with Wi-Fi pan/tilt control~(see \autoref{fig: hardware}~\circled{5}). 
To be noted, this head-mounted solution for projector placement was determined for two key reasons: (1) unlike an arm-mounted solution, it does not interfere with the robot's ability to perform gestures; (2) it provides a full 360-degree projection range, overcoming the body occlusion that prevents a torso-mounted projector from reaching the robot's backside.

The mobile chassis is a Wheeltec S300\footnote{\url{https://wheeltec.net/}} with dual M10P LiDARs for 360° Simultaneous Localization and Mapping (SLAM), controlled via a Jetson Orin Nano~(see \autoref{fig: hardware}~\circled{1}). The robot's shell is 3D-printed in PLA~(see \autoref{fig: hardware}~\circled{7}), and its face is made of semi-transparent PETG housing an LED breathing light to indicate speaking status~(see \autoref{fig: hardware}~\circled{6}). The projector and 2-DOF gimbal communicate with the control PC via MQTT\footnote{Message Queuing Telemetry Transport: a lightweight publish/subscribe messaging protocol widely used in IoT/robots.} over Wi-Fi, while the Gesture Actuation Module and mobile base use CAN bus\footnote{Controller Area Network: a robust, low-level communication network originally from automotive systems, used for reliable device control.} with Robot Operating System (ROS) topics for control. During guided tours, the robot follows SLAM-based maps with preset device positions, with occasional human intervention for minor adjustments due to navigation limits.


\begin{figure*}[tb]
  \centering
  \includegraphics[width=\textwidth]{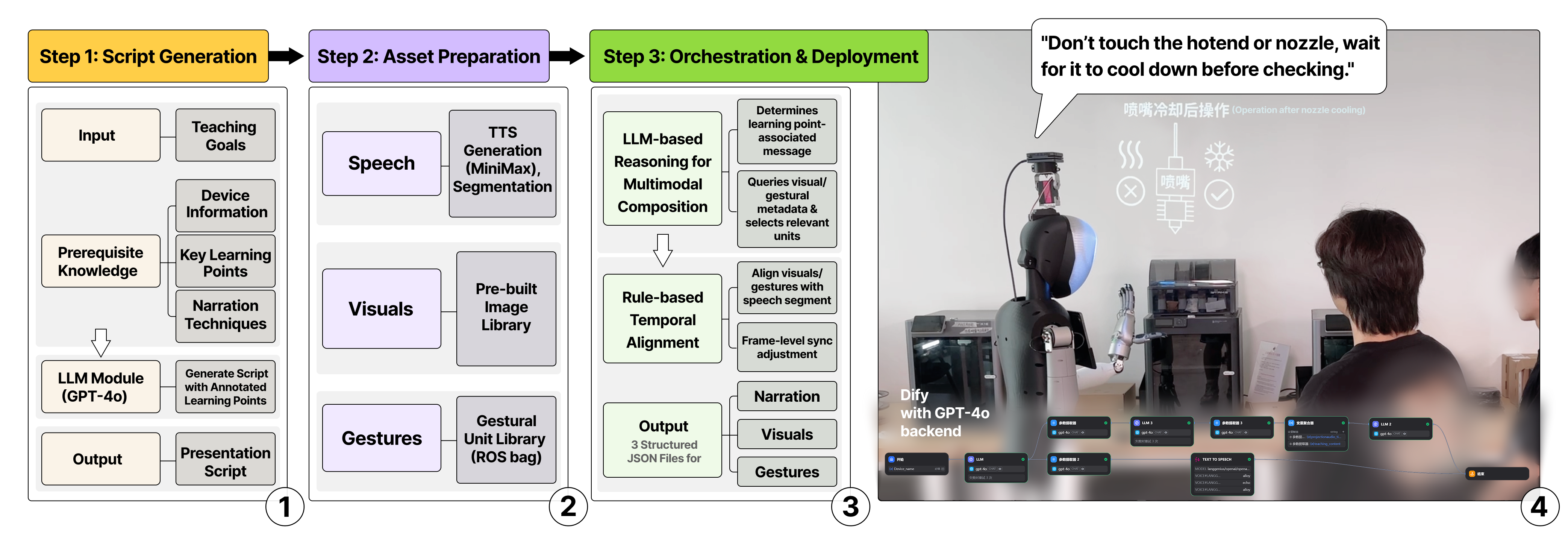}
  \caption{Presentation choreography workflow, aligning speech, visuals, and gestures on a unified timeline.}
  \Description{A flowchart detailing a presentation choreography workflow, organized horizontally across four main stages. Stages 1-3 contain 26 primary elements connected by flow arrows, starting with 'Input' and ending with a final 'Output' of structured files. Stage 4 is a photograph illustrating the workflow's application.
The workflow details are as follows:
Stage 1: Script Generation.
  1. The process starts with two main inputs: 'Input' (defined by 'Teaching Goals') and 'Prerequisite Knowledge' (defined by 'Device Information', 'Key Learning Points' and 'Narration Techniques').
  2. These inputs feed into the LLM (Large Language Model) Module (GPT-4o).
  3. This module's task is to 'Generate Script with Annotated Learning Points.
  4. The final result of this stage is an 'Output,' which is the 'Presentation Script.' This script flows to Stage 2 and 3.
Stage 2: Asset Preparation.
  This stage shows three parallel tracks for creating assets:
  1. Speech: Processed via TTS (Text-to-Speech) Generation (MiniMax), Segmentation.
  2. Visuals: Sourced from a Pre-built Image Library.
  3. Gestures: Sourced from a Gestural Unit Library (ROS bag).
Stage 3: Orchestration & Deployment.
  1. Assets from Stage 2 and learning points from Stage 1 flow into LLM-based Reasoning for Multimodal Composition. This element 'Determines learning point-associated message' and 'Queries visual/gestural metadata & selects relevant units.'
  2. The output flows to a Rule-based Temporal Alignment module, which performs 'Align visuals/gestures with speech segment' and 'Frame-level sync adjustment.'
  3. The final output is 3 Structured JSON (JavaScript Object Notation) Files for 'Narration,' 'Visuals,' and 'Gestures.'
Stage 4: Practical Example.
  This is a photograph showing the deployed result. The robot delivers a verbal warning through a speech bubble (“Don’t touch the hotend or nozzle...”), performs a blocking hand gesture, and simultaneously projects a visual safety warning on the wall that illustrates the nozzle should only be operated after cooling down. An overlay on the image displays the Dify backend interface, which visualizes the workflow as a node-based graph.}
  \label{fig: choreography_workflow}
\end{figure*}

\subsubsection{\textbf{Presentation Choreography Workflow}}
This workflow enables educators and technicians to edit, preview, and verify ProjecTA's presentations across modalities.
Based on input teaching goals, i.e., which equipment to cover and in what sequence, it compiles an executable choreography script that controls ProjecTA during a tour. 
The workflow assembles and schedules (1) generated speech segments, (2) existing visual display assets, and (3) predefined gesture units, then coordinates them across channels at runtime.
The workflow also makes ProjecTA readily extensible and deployable in future similar learning contexts (other makerspaces or museums, etc.).

\textbf{\textit{Presentation Script Generation}}: As shown in \autoref{fig: choreography_workflow}~\circled{1}, this LLM-based module generates ProjecTA's script based on the teaching goals (which equipment to cover and in what sequence) and a set of prerequisite knowledge.

\textbf{[Prerequisite Knowledge]} consists of Device Information, Key Learning Points, and Narration Techniques.

\textit{Device Information.} The descriptive information about all pieces of equipment in the makerspace was pre-collected from reliable sources.

\textit{Key Learning Points.} Based on experts' verbalization in the bodystorming session, we compiled the essential learning points that should be covered when explaining each piece of makerspace equipment to beginners. These points include basic knowledge about how it works, how to operate it, and safety precautions \textbf{(DR1)}, e.g., ``PLA material is recommended to be printed at 210–220 °C'' (for more details, please refer to Supplementary Material A). We also referenced resources widely used in maker education to ensure comprehensiveness \cite{barbara2024using,love2018perceptions,lundberg2018foundational}. 
These resulting key learning points serve as the central materials: ProjecTA's presentation scripts are structured around them, and its visual assets and gestural units are also designed to communicate these key learning points effectively to novice learners.

\textit{Narration Techniques.} Narration techniques are extracted from the bodystorming enactments by the experts, e.g., using everyday analogies to facilitate comprehension, or inserting transitions between, or drawing connections across related machines \textbf{(DR2)}. Based on both expert enactments and our own testing, the narration for each device was kept to about 4–5 minutes.

\textbf{[Script Generation]}: We used the \textit{GPT-4o-2024‑08‑06}\footnote{\url{https://platform.openai.com/}} model to generate the presentation scripts. The aforementioned \textit{Device Information}, \textit{Key Learning Points}, and \textit{Narration Techniques} were integrated into the instruction prompt (see Supplementary Material A for the prompt). The prompt required the LLM to explicitly mark where each \textbf{key learning point} appears in the generated script, enabling easier verification and supporting subsequent multimodal orchestration.

\textbf{\textit{Preparation of Speech, Visual Assets, and Gestural Units}}: As shown in \autoref{fig: choreography_workflow}~\circled{2}, the preparation of multimodal resources are outlined below:

\textbf{[Speech Generation and Segmentation]}: The generated script was passed to MiniMax's text-to-speech (TTS) service\footnote{\url{https://www.minimaxi.com/}}. As voice timbre was not a focus of this study, we adopted a standard explanatory voice from the MiniMax platform (see supplementary video). The generated audio was segmented into units according to semantic boundaries, ensuring each segment contained at most one \textit{key learning point} for scheduling. Metadata for each speech segment included its duration and the \textit{key learning point} it was aligned with.

\begin{figure*}[tb]
  \centering
  \includegraphics[width=\textwidth]{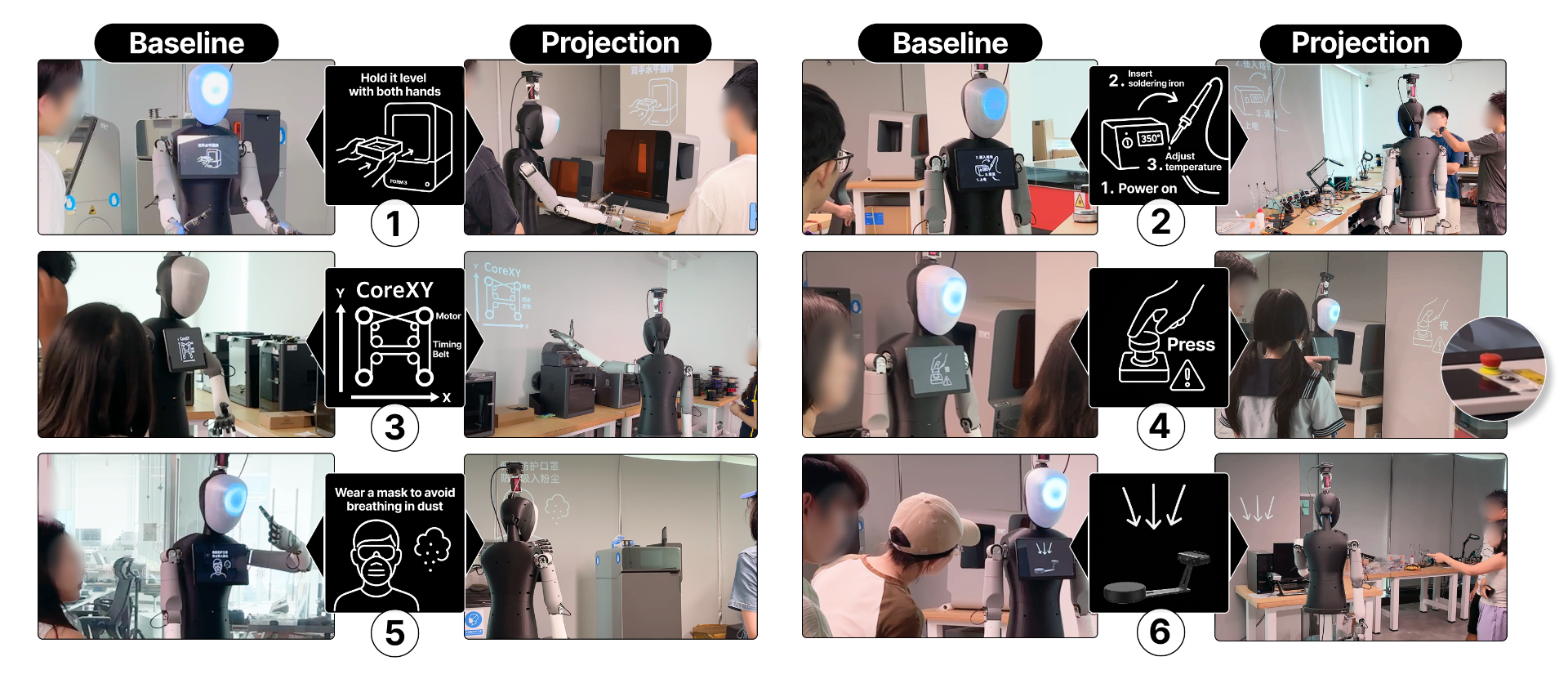}
  \caption{Comparative examples of Baseline (chest-mounted screen) vs. ProjecTA (in-situ projection) guidance in makerspace tours.}
  \Description{A figure presenting six numbered side-by-side comparisons between two robot guidance methods: "Baseline" and "Projection." In each comparison, the "Baseline" image on the left shows a humanoid robot displaying instructions on its chest-mounted screen. The corresponding "Projection" image on the right shows the same robot delivering the same instruction by projecting visuals directly onto a wall or near the relevant physical object. The six scenarios depicted are: (1) holding a component level, (2) operating a soldering iron, (3) explaining a CoreXY mechanism, (4) pressing a emergency stop button, (5) wearing a mask, and (6) providing a directional cue for a 3D scanner.}
  \label{fig: baseline_comparison}
\end{figure*}

\textbf{[Visual Display Assets Library]}: The visual display assets were organized as a pre-built image library designed to support novices to understand and consolidate each \textit{key learning point}. Note that some \textit{key learning points} require multiple images when they span steps of sub-concepts. As shown in \autoref{fig: baseline_comparison}~\circled{5}, each image uses simplistic, clear graphics with minimal labels aligned with the narration \textbf{(DR5)}. 

The metadata includes each visual display asset's intended projection location in the makerspace (e.g., on the equipment or nearby surfaces), a textual description, and the linked \textit{key learning point} for later orchestration. See Supplementary Material B for details.

\textbf{[Robot's Gestural Unit Library]}: We constructed a library of 42 robot gesture-control sequences, derived from 45 video-recorded bodily enactments that the experts performed to explain specific \textit{key learning points}. These gestures included deictic \textbf{(DR3)}, iconic, and metaphoric forms \textbf{(DR4)}. Highly similar examples were merged, and each remaining sequence was teleoperated and recorded as a ROS bag file, producing a discrete gestural unit that can be triggered via the \textbf{Gesture Actuation Module}. Note that a single \textit{key learning point} may map to multiple gestural units to support flexible orchestration.

The metadata of each unit includes the robot's spatial location and orientation, matching the expert's original demonstration for the given equipment, a textual description, a contextual note (indicating which equipment and \textit{key learning point} it addresses and the accompanying narration), and its duration. See Supplementary Material C for details.

\textbf{\textit{Orchestration of Speech, Visual Assets, and Gestural Units}}: As shown in \autoref{fig: choreography_workflow}~\circled{3}, the orchestration process has two steps:

\textbf{[LLM-based Reasoning for Multimodal Composition]}: For each \textbf{speech segment}, the system determines whether its learning point-associated message needs further illustration or reinforcement, and then queries visual and gestural library metadata to select and combine the most relevant images, gestural units, or both.

\textbf{[Rule-based Temporal Alignment]}: Selected visuals and gestures are then aligned temporally with the associated speech segment \textbf{(DR6)}. 
Selected display assets remain visible for the full segment; gestures are synchronized to playback.
When multiple images are needed, each is aligned to the exact narration moment it explains (frame-level synchronization).
If a gesture exceeds the segment length, the inter-segment pause is extended to prevent overlap.
Educators and technicians can preview the outcome on the robot and fine-tune timings across the modalities.

In our system, multimodal resources were orchestrated along a unified timeline. The LLM-based agentic pipeline was constructed upon Dify\footnote{\url{https://dify.ai/}}, with \textit{GPT-4o-2024‑08‑06} as the base model (refer \autoref{fig: choreography_workflow}~\circled{4}) to achieve multimodal orchestration. All reasoning is performed offline to avoid runtime latency and allow for human preview and verification. The system outputs three structured JSON files encompassing narration, visuals, and gestures, each with execution times; the backend server converts these into synchronized, executable robot behaviors. This choreography system also enables users to pre-generate multiple narration variants and select the preferred one for further edits or final deployment.

\section{Methodology}
Our study aims to address the follow question: How does a robotic TA with in-situ projection, compared with a screen-based counterpart, affect learners’ experiences during makerspace tours? \textbf{(RQ)} To investigate this question, we built and deployed ProjecTA in a real university makerspace (\autoref{fig: realmakerspace}). Adopting a mixed-methods, within-subject comparative approach, ProjecTA was contrasted with Baseline, a functionally equivalent counterpart using the chest-screen as its visual display.

\subsection{Baseline system for Comparison}
For comparison, the Baseline and ProjecTA systems were deployed on identical robotic hardware equipped with both a chest-mounted screen (\autoref{fig: hardware}~\circled{8}) and a head-mounted projector (\autoref{fig: hardware}~\circled{3}). This configuration extends prior findings of spatial AR and HUD studies in stationary operation (e.g., assembling or driving) to nomadic learning with a mobile robot. In our study, the guided walk-and-talk tour serves as a typical nomadic learning setting, providing a baseline cognitive-load context for comparing the two display modalities. The Baseline condition replicated the standard screen-based displays of existing robotic platforms like Pepper \cite{pandey2018mass}. Conversely, the ProjecTA condition utilized in-situ projection to examine differences in learner experiences, such as extraneous cognitive load.
To ensure a fair comparison, both conditions ran on the same choreography instance generated by the Presentation Choreography Workflow, including the same narration, gestures, visual assets, tour route, and timing.
To further ensure fairness for the Baseline, we made two necessary screen-specific optimizations: (1) deictic gestures that, in ProjecTA, pointed to projected overlays were replaced with gestures cueing viewers to look at the screen (\autoref{fig: baseline_comparison}~\circled{3}); (2) spatial markers that ProjecTA placed on or around the physical equipment (arrows, halos, highlights) were mirrored on the screen by showing an image of the same equipment with identical markers (see \autoref{fig: baseline_comparison}~\circled{6}). 
All other assets, including movement animations and learning point illustrations, were unchanged (see \autoref{fig: baseline_comparison}~\circled{1}\circled{2}\circled{4}\circled{5}). 
Therefore, the sole significant difference between the two conditions was the display modality (on-body screen versus in-situ projection), allowing us to assess their effects on learners' cognitive load and experiences during makerspace tours.

\subsection{Participants}
This study recruited 24 participants (13 male, 11 female; age 20–31, M = 24.25, SD = 2.61) via social-meadia posts. We targeted novice learners and screened out registrants who reported with rich makerspace experience before scheduling. We collected demographics via an online questionnaire, and respondents included university students and working professionals, spanning backgrounds in engineering/sciences (n = 14), humanities (n = 3), social sciences (n = 4), and art/design (n = 3). Sessions were conducted in dyads (12 pairs): 6 pairs who knew each other and 6 pairs of strangers. We assessed two aspects of participants' past experiences using five-point rating scales (1 = never, 5 = often): hands-on activities in makerspaces (19/24 rated 1) and guided equipment tours (22/24 rated 1). Familiarity with makerspace equipment was also measured on a five-point rating scale (1 = not at all familiar, 5 = very familiar), and participants mostly reported low familiarity. Detailed familiarity ratings are provided in \autoref{Appendix: familiarity}.

\begin{figure*}[tb]
  \centering
  \includegraphics[width=\textwidth]{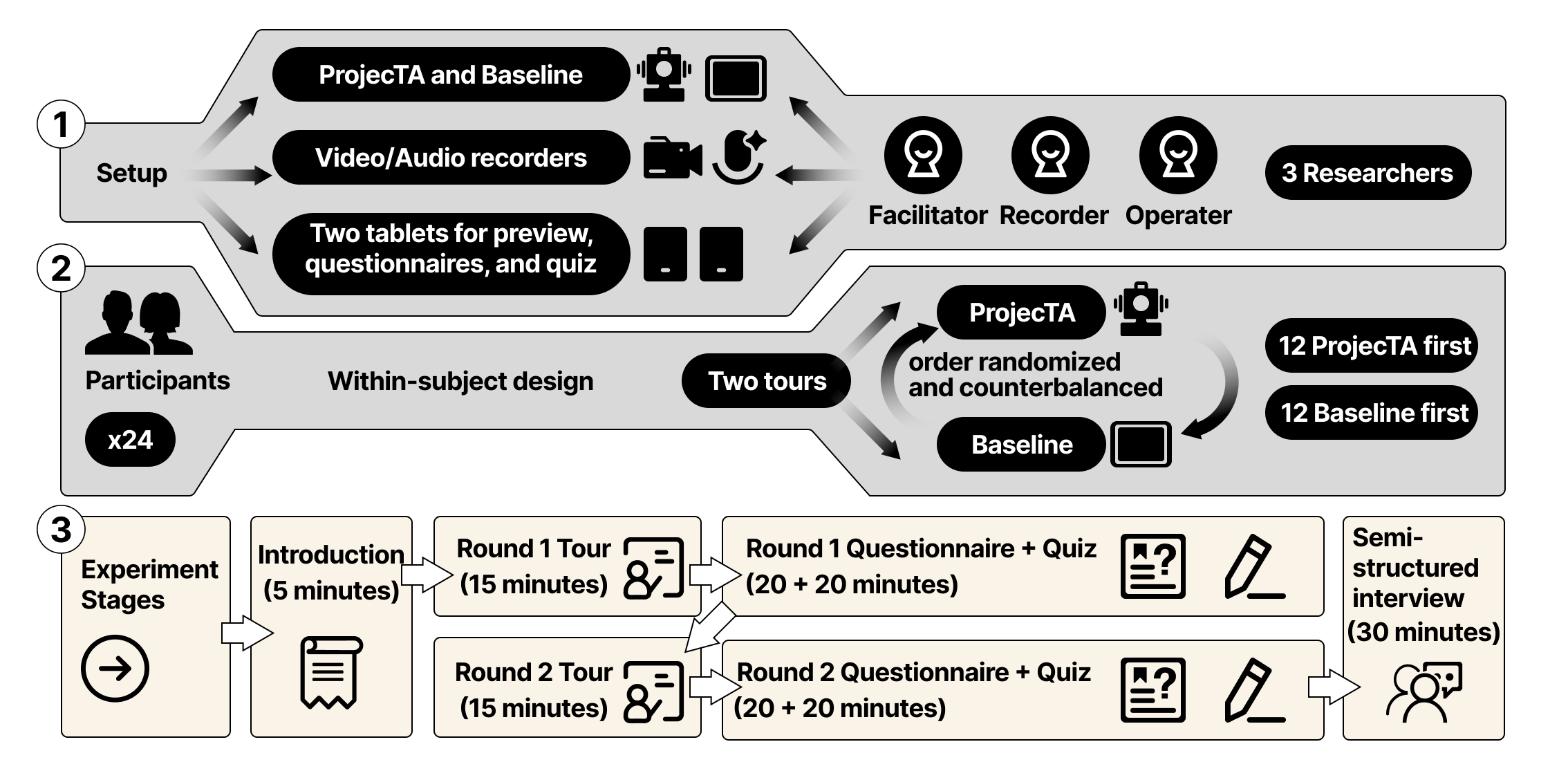}
  \caption{\protect\circled{1} Experimental setup, \protect\circled{2} experimental design and \protect\circled{3} experiment stages.}
  \Description{A flowchart in three numbered horizontal sections detailing the user study procedure.
  Section 1, "Setup," shows three outputs: "ProjecTA and Baseline," "Video/Audio recorders," and "Two tablets." These elements receive inputs from "3 Researchers," who are specified as "Facilitator," "Recorder," and "Operater."
  Section 2, "Participants," starts with "24 Participants." It flows to "Within-subject design" and then to "Two tours." From "Two tours," a counterbalanced flow splits into "ProjecTA" and "Baseline," which then leads to two final groups: "12 ProjecTA first" and "12 Baseline first."
  Section 3, "Experiment Stages," is a linear sequence. The start state is "Introduction (5 minutes)." It flows to two parallel rounds. Round 1 consists of "Round 1 Tour (15 minutes)" which flows to "Round 1 Questionnaire + Quiz (20 + 20 minutes)." Round 2 consists of "Round 2 Tour (15 minutes)" which flows to "Round 2 Questionnaire + Quiz (20 + 20 minutes)." The outputs of both rounds converge into the end state, "Semi-structured interview (30 minutes)."}
  \label{fig: setup}
\end{figure*}

\subsection{Setup and Procedure}
\begin{figure*}[tb]
  \centering
  \includegraphics[width=\textwidth]{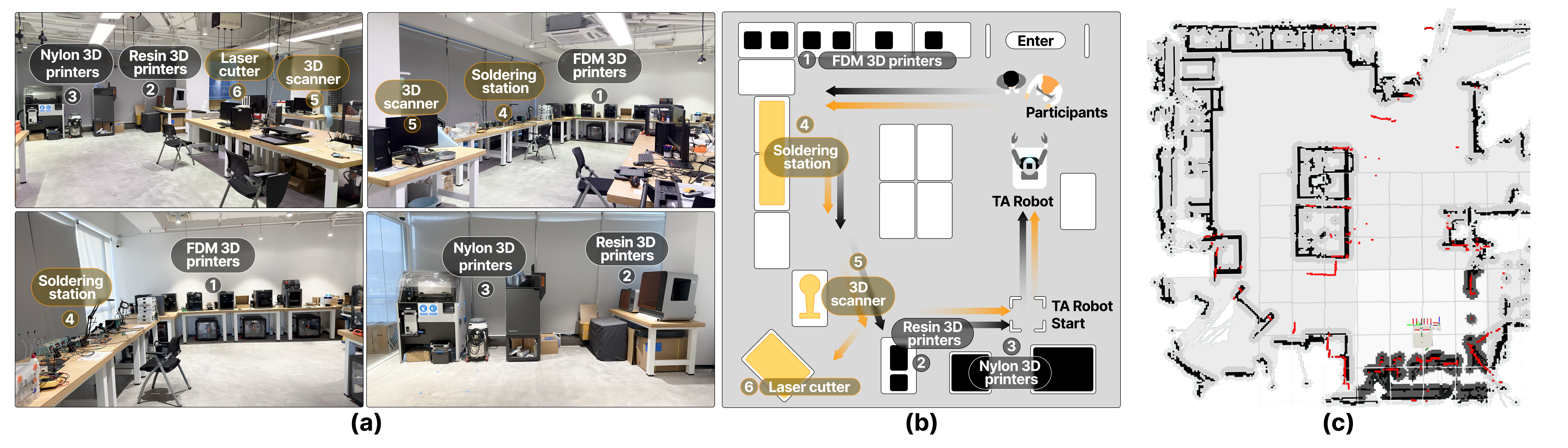}
  \caption{Experimental environment: (a) photos of the university makerspace, (b) top-down schematic of equipment layout and robot path, and (c) RViz map of the space.}
  \Description{A three-part figure detailing the experimental makerspace environment.
  Part (a) shows four photographs of the physical space in real makerspace with six key equipment areas labeled numerically: (1) FDM 3D printers, (2) Resin 3D printers, (3) Nylon 3D printers, (4) Soldering station, (5) 3D scanner, and (6) Laser cutter.
  Part (b) is a top-down schematic showing the guided tour path. The flow starts with the TA Robot and Participants at the 'Start' point. It proceeds via a sequence of arrows to the six numbered stations from Part (a), beginning at station 1 and concluding at station 6.
  Part (c) is a 2D occupancy grid map from RViz. It displays the layout with detected obstacles and walls in black, navigable free space in grey, unknown areas in white, and laser scan sensor readings as red dots.}
  \label{fig: realmakerspace}
\end{figure*}

The experiment was conducted in a university makerspace (\autoref{fig: realmakerspace}). As shown in \autoref{fig: setup}~\circled{1}, the setup consisted ProjecTA and Baseline; video and audio recorders for data collection; and two additional tablets for participants to preview the procedure and complete questionnaires and post-tour quiz during the session. Two researchers were present in each session: one facilitated the session and one recorded the procedure. A third researcher, located in an adjacent space, operated the backend system.


To compare ProjecTA with the Baseline, we used a within-subject comparative design (see \autoref{fig: setup}~\circled{2}). Each participant dyad completed two robot-guided tours, one with ProjecTA and one with the Baseline; modality order was randomized and balanced (12 ProjecTA first, 12 Baseline first). To avoid learning the same content twice, we prepared two non-overlapping tour sets, A and B, each covering three different machines: Set A included the FDM 3D printer (\autoref{fig: realmakerspace}~\circled{1}), Resin 3D printer (\autoref{fig: realmakerspace}~\circled{2}), and Nylon 3D printer (\autoref{fig: realmakerspace}~\circled{3}); Set B included the Soldering station (\autoref{fig: realmakerspace}~\circled{4}), 3D scanner (\autoref{fig: realmakerspace}~\circled{5}), and Laser cutter (\autoref{fig: realmakerspace}~\circled{6}). The tour sets were treated as a counterbalanced random factor to offset learning effects. Every participant experienced both sets across the two rounds, with set order counterbalanced within each modality order: half of the participants experienced A first and half B first. This scheme isolates the effect of display modality while mitigating cross-condition learning effects \cite{bartels2010practice} and order effects. The experimental stages (\autoref{fig: setup}~\circled{3}) were as follows:


\textbf{Introduction} (5 minutes): The researcher introduced the overall procedure, including the structure of the two tour-guided rounds and emphasized safety precautions (e.g., the robot's emergency stop). Participants then signed consent forms and completed a demographic questionnaire, along with questions about prior experience with makerspace equipment and guided tours.

\textbf{Round 1 Tour} (15 minutes): Participants completed the first round of robot-guided learning with one display modality (projector or screen). Two participants jointly followed the explanations.

\textbf{Round 1 Questionnaire + Quiz} (20 + 20 minutes): After the tour, participants first completed questionnaires that included measures of cognitive load, user engagement, and a customized scale, followed by a first-round-specific post-tour quiz aligned with three devices introduced in that round.

\textbf{Round 2 Tour} (15 minutes): Participants experienced the other display modality in the second round. The robot introduced the remaining three devices. The process was identical to Round 1.

\textbf{Round 2 Questionnaire + Quiz} (20 + 20 minutes): As in Round 1, participants completed a same questionnaire and a second-round-specific post-tour quiz.

\textbf{Semi-structured interview} (30 minutes): After both rounds, participants engaged in a semi-structured interview and participants received 200 CNY as compensation for their time.

\subsection{Data Gathering}

\subsubsection{\textbf{Questionnaire}}

\begin{table*}[t]
\centering
\footnotesize
\begin{tabular}{llp{7cm}}
\toprule
\textbf{ID} & \textbf{Caption} & \textbf{Details} \\
\midrule
RS1--4   & \textbf{Intrinsic Load} (learning material's inherent complexity)      & the Intrinsic Load subscale in Cognitive Load Scale (CLS)\cite{leppink2015evolution} \\
\cmidrule(lr){1-3}
RS5--8   & \textbf{Extraneous Load} (avoidable effort caused by presentation)    & the Extraneous Load subscale in Cognitive Load Scale (CLS)\cite{leppink2015evolution} \\
\cmidrule(lr){1-3}
RS9--11  & \textbf{Focused Attention} (immersiveness and absorption)       & the FA subscale in User Engagement Scale Short Form (UES-SF)~\cite{o2018practical} \\
\cmidrule(lr){1-3}
RS12--14 & \textbf{Perceived Usability}           & the PU sbscale in User Engagement Scale Short Form (UES-SF)~\cite{o2018practical} \\
\cmidrule(lr){1-3}
RS15--17 & \textbf{Aesthetic Appeal}          & the AE subscale in User Engagement Scale Short Form (UES-SF)~\cite{o2018practical} \\
\cmidrule(lr){1-3}
RS18--20 & \textbf{Reward Factor} (rewarding experience)             & the RW subscale in User Engagement Scale Short Form (UES-SF)~\cite{o2018practical} \\
\cmidrule(lr){1-3}
RS21     & \multirow{2}{*}{\centering\arraybackslash \textbf{Usefulness of visual display}} 
         & The visual display of the robot system was very helpful for my understanding of the presentation content. \\
\cmidrule(lr){1-1}\cmidrule(lr){3-3} 
RS22     &                     & The visual display of the robot system made it easier for me to focus my attention on the corresponding area of the actual equipment or space. \\
\cmidrule(lr){1-3}
RS23     & \multirow{3}{*}{\centering\arraybackslash \textbf{Multi-modal complementary} (visuals, gestures, and speech)} 
         & When the robot system performed an action, I could easily locate the related information on its visual display. \\
\cmidrule(lr){1-1}\cmidrule(lr){3-3} 
RS24     &                     & The robot system's actions, speech, and visual display content were well-coordinated. \\
\cmidrule(lr){1-1}\cmidrule(lr){3-3} 
RS25     &                     & The visual display of the robot system effectively complemented its speech and actions. \\
\bottomrule
\end{tabular}
\caption{Item-type distribution in rating scales.}
\Description{A table detailing the rating scales and questionnaire items used in the study. The table is organized into three columns: "ID," "Caption," and "Details." The rows define the constructs being measured, such as "Intrinsic Load" and "Perceived Usability," by providing their item identifiers, names, and a detailed description or the source of the scale.}
\label{tab:rating-scales}
\end{table*}

The questionnaire comprised 25 rating items (RS1–RS25; \autoref{tab:rating-scales}).
RS1–RS8 used the streamlined Cognitive Load Scale (CLS)~\cite{leppink2015evolution} with two subscales: Intrinsic Load (RS1–RS4), capturing the effort caused by the learning materials' inherent complexity and learners prior knowledge, and Extraneous Load (RS5–RS8), capturing the avoidable effort induced by the presentation of the learning materials. 
Because the equipment sets (learning materials) were counterbalanced within each condition, we did not expect condition differences on Intrinsic Load. 
Our primary test was whether presentation mode (ProjecTA vs. the Baseline condition) differentially affected Extraneous Load.

RS9–RS20 used the User Engagement Scale Short Form (UES-SF)~\cite{o2018practical} with four subscales: Focused Attention (RS9–RS11; experienced immersiveness or absorption), Perceived Usability (RS12–RS14), Aesthetic Appeal (RS15–RS17), and Reward Factor (RS18–RS20). 
We used the UES-SF to assess experiential differences beyond cognitive load.

RS21–RS25 were custom items aligned with our research goals. RS21–RS22 assessed the perceived usefulness of the visual displays for (a) understanding the presented content and (b) maintaining focus on the relevant object or region in the space. 
RS23–RS25 assessed perceived multimodal complementarity among visual display, gesture, and speech across the two conditions.

\subsubsection{\textbf{Post-tour Quiz}}



Each tour was followed by a post-tour quiz. Each quiz comprised 30 items covering all three pieces of equipment from that tour. Items targeted each equipment's \textit{Learning Points}: how it works, how to operate it, and safety precautions. 
Question formats included single-answer multiple-choice and ordering (sequence) items. 
Each correct response (i.e., a correct option or a correctly placed position in an ordering item) was worth one point; raw scores were then linearly normalized to a 0–100 scale for analysis. 
After each tour, participants completed the questionnaire first and then the tour-specific quiz. This was intended to insert a brief, content-related delay, reducing reliance on immediate/short-term memory (e.g., verbatim recall of just-seen labels) and yielding a more valid assessment.

\subsubsection{\textbf{Interview}}
We conducted semi-structured interviews in dyads to collect in-depth user feedback. Interviews began with questions about participants' experiences in both conditions, probing how they followed the presentations in each round and what difficulties they encountered. We then asked them to describe positive and negative experiences with each condition and to suggest improvements. Next, participants were asked to provide concrete examples from both conditions and state their preference with rationales. Finally, we invited suggestions for other scenarios where the system could be applied. 
Each interview lasted around 20 minutes and was video-recorded for analysis; the complete question list is provided in \autoref{Semi-structured}. 
We conducted a thematic analysis following Blandford’s guidelines~\cite{blandford2016qualitative}, two researchers conducted collaborative inductive coding. They initially annotated the transcript to identify relevant quotes, key concepts, and preliminary patterns in the data. These initial insights were further developed through regular discussions among four researchers, leading to a detailed coding scheme aligned with the research objectives. Quotes were then coded and clustered into a hierarchy of emerging themes, continually reviewed, and refined in recurrent meetings, where exemplar quotes were also selected to illustrate each theme and sub-theme. 
Alongside this, the team reviewed and annotated the session videos, keeping both the research questions and the emerging thematic structure in view. We collected the video segments that served as evidence or exemplars for the thematic analysis results, especially those highlighting behaviors of participants during nomadic guiding tour in makerspace. In addition, we supplemented our analysis with photographic documentation of key interaction moments, spatial arrangements, and notable projection–object–participant configurations observed during the sessions.

\section{Findings}

\begin{figure*}[tb]
  \centering
  \includegraphics[width=\textwidth]{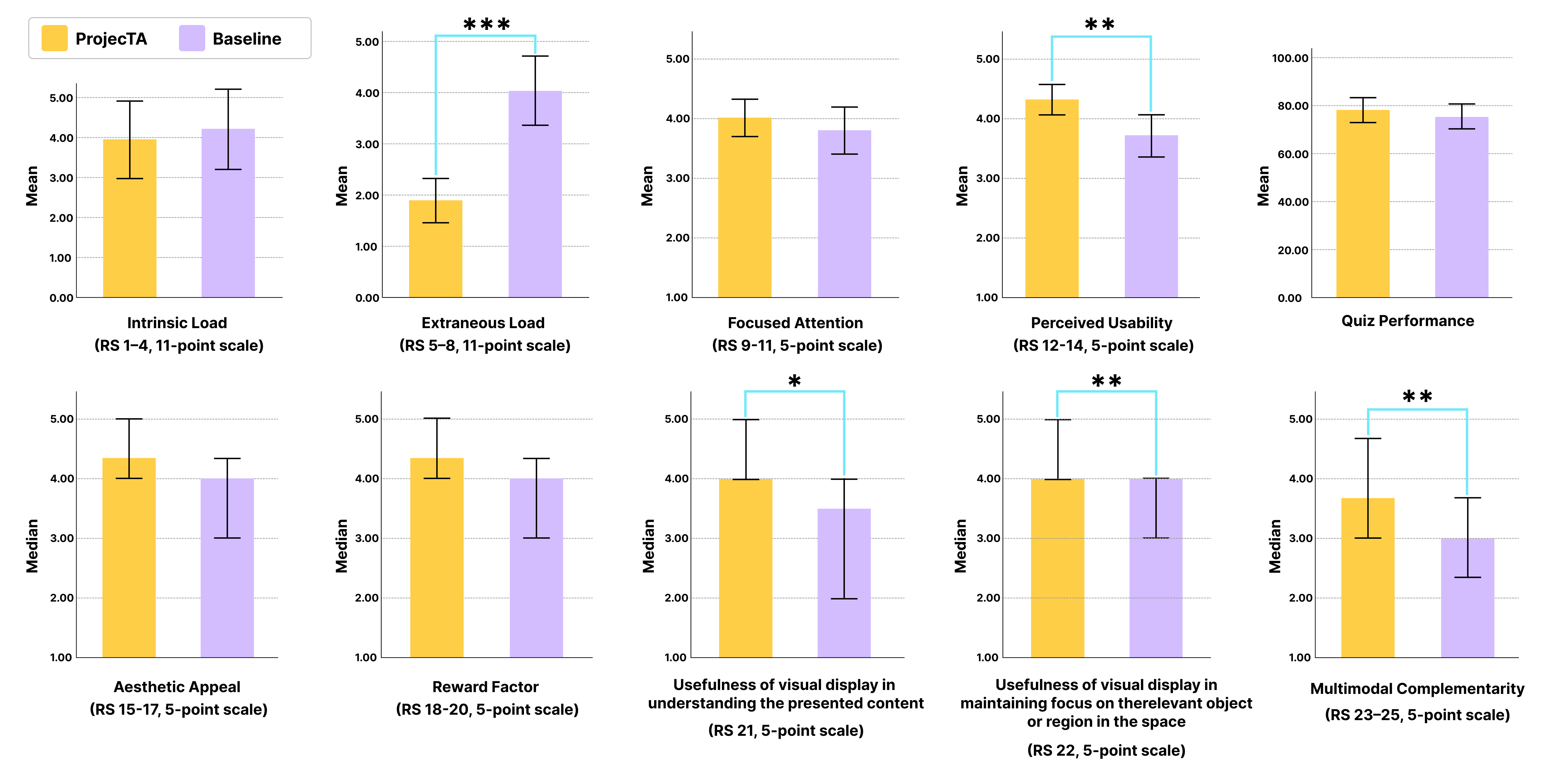}
  \caption{Quantitative results comparing ProjecTA with Baseline (*p < .050, **p < .010, ***p < .001; error bars indicate 95\% CI).}
  \Description{A set of ten bar charts comparing user study results for the ProjecTA and Baseline conditions across different metrics. The figure is arranged in a two-by-five grid. Each individual chart contains a pair of vertical bars with error bars: a yellow bar representing ProjecTA and a light purple bar representing Baseline. The vertical axis of the top row of five charts represents the mean scores, while the vertical axis of the bottom row of five charts represents the median scores. Cyan brackets with one, two, or three asterisks are used to visually highlight pairs with statistically significant differences.Stars next to the values indicate levels of statistical significance: one star means p < .050, two stars mean p < .010, and three stars mean p < .001}
  \label{fig: quantitative result}
\end{figure*}

\subsection{Quantitative Results}
The internal consistency of two custom subscales was assessed. Cronbach's $\alpha$ indicated acceptable reliability for Multimodal Complementarity (RS23–RS25) in both conditions (ProjecTA $\alpha$ = .75; Baseline $\alpha$ = .72; > .70). Visual Display Usefulness items (RS21–RS22) were < .70 in both conditions and were analyzed separately.

Normality of within-participant difference scores was tested using the Shapiro–Wilk test. Assumptions were met for Intrinsic Load (RS1–RS4), Extraneous Load (RS5–RS8), Focused Attention (RS9–RS11), Perceived Usability (RS12–RS14), and Quiz Performance; these outcomes were compared with paired-samples t-tests. 
Aesthetic Appeal (RS15–RS17), Reward Factor (RS18–RS20), RS21, RS22, and Multimodal Complementarity (RS23–RS25) violated normality and were analyzed with Wilcoxon signed-rank tests, The results are presented in \autoref{fig: quantitative result}.

\textbf{Intrinsic Load (RS1-4)} captures the effort required by the material's inherent complexity and the learner's prior knowledge. No significant difference was found: $M_{\text{ProjecTA}}=3.96$, $SD_{\text{ProjecTA}}=2.31$; $M_{\text{Baseline}}=4.21$, $SD_{\text{Baseline}}=2.36$; $M_{\text{diff}}=-0.25$, $SD_{\text{diff}}=2.50$; $t(23)=-0.489$, $p=0.629$, Cohen's $d=-0.100$.

\textbf{Extraneous Load (RS5-8)} captures the avoidable effort introduced by how information is presented. \textit{ProjecTA} was significantly lower than \textit{Baseline} with a large effect size (Cohen's $|d| > 1$): $M_{\text{ProjecTA}}=1.90$, $SD_{\text{ProjecTA}}=1.01$; $M_{\text{Baseline}}=4.04$, $SD_{\text{Baseline}}=1.61$; $M_{\text{diff}}=-2.15$, $SD_{\text{diff}}=1.98$; $t(23)=-5.322$, $p<0.001$, $d=-1.086$.

\textbf{Focused Attention (RS9-11)} captures immersiveness and absorption in experience. No significant difference found: $M_{\text{ProjecTA}}=4.00$, $SD_{\text{ProjecTA}}=0.73$; $M_{\text{Baseline}}=3.79$, $SD_{\text{Baseline}}=0.93$; $M_{\text{diff}}=0.21$, $SD_{\text{diff}}=0.73$; $t(23)=1.390$, $p=0.178$, $d=0.284$.

\textbf{Perceived Usability (RS12-14)} were reverse-worded and were reverse-scored for analysis and presentation (higher scores reflect greater usability). \textit{ProjecTA} was significantly higher than \textit{Baseline}: $M_{\text{ProjecTA}}=4.31$, $SD_{\text{ProjecTA}}=0.60$; $M_{\text{Baseline}}=3.71$, $SD_{\text{Baseline}}=0.84$; $M_{\text{diff}}=0.60$, $SD_{\text{diff}}=0.86$; $t(23)=3.393$, $p<0.01$, $d=0.692$.

\textbf{Quiz Performance}. No significant difference found: $M_{\text{ProjecTA}}=78.13$, $SD_{\text{ProjecTA}}=12.11$; $M_{\text{Baseline}}=75.52$, $SD_{\text{Baseline}}=12.16$; $M_{\text{diff}}=2.61$, $SD_{\text{diff}}=13.77$; $t(23)=0.927$, $p=0.364$, $d=0.189$.

\textbf{Aesthetic Appeal (RS15–17)}. No significant difference (Wilcoxon): $Z=-1.371$, $p=0.171$, $r=0.280$. Medians (IQR): \textit{ProjecTA} $=3.17$ [$2.67$, $4.00$]; \textit{Baseline} $=3.00$ [$2.42$, $3.92$]; median difference $=0.00$ [$-0.33$, $0.33$].

\textbf{Reward Factor (RS18–20)}. No significant difference (Wilcoxon): $Z=-1.703$, $p=0.089$, $r=0.348$. Medians (IQR): \textit{ProjecTA} $=4.33$ [$3.75$, $5.00$]; \textit{Baseline} $=4.00$ [$2.75$, $4.83$]; median difference $=0.17$ [$0.00$, $0.67$].

\textbf{Usefulness of visual display in understanding the presented content (RS21)} \textit{ProjecTA} significantly exceeded \textit{Baseline} with a medium-large effect (Wilcoxon): $Z=-2.429$, $p<0.05$, $r=0.496$. Medians (IQR): \textit{ProjecTA} $=4.00$ [$4.00$, $5.00$]; \textit{Baseline} $=4.00$ [$3.00$, $4.00$]; median difference $=0.00$ [$0.00$, $1.00$].

\textbf{Usefulness of visual display in maintaining focus on the relevant object or region in the space (RS22)} \textit{ProjecTA} significantly exceeded \textit{Baseline} with a large effect (Wilcoxon): $Z=-3.096$, $p<0.01$, $r=0.632$. Medians (IQR): \textit{ProjecTA} $=4.00$ [$4.00$, $5.00$]; \textit{Baseline} $=3.50$ [$2.00$, $4.00$]; median difference $=1.00$ [$0.00$, $2.00$].

\textbf{Multimodal Complementarity (RS23-25)} assesses the perceived coordination and mutual reinforcement among the robot's visual displays, gestures, and speech. \textit{ProjecTA} significantly exceeded \textit{Baseline} with a large effect (Wilcoxon): $Z=-2.958$, $p<0.01$, $r=0.604$. Medians (IQR): \textit{ProjecTA} $=3.67$ [$3.00$, $4.67$]; \textit{Baseline} $=3.00$ [$2.33$, $3.92$]; median difference $=0.50$ [$0.00$, $1.00$].

Overall, the most significant finding is that ProjecTA substantially reduced extraneous cognitive load: the avoidable effort demanded by the presentation of learning materials \cite{ChandlerSweller1991}. While both the ProjecTA condition and baseline condition achieved comparable quiz scores, ProjecTA's ability to lessen avoidable cognitive effort suggests strong potential for future robotic TA implementations.
As expected from the counterbalanced design, there was no significant difference in the learning materials' inherent difficulty (intrinsic load). 
Because the two conditions used identical hardware and the same choreography and differed only in visual modality, there were no significant differences in aesthetic appeal, immersiveness (focused attention), or rewarding experience. By contrast, ProjecTA scored significantly higher on outcomes tied to information delivery and cross-modal coordination: higher perceived usability, more effective visual displays for understanding content and staying oriented to physical referents, and stronger multimodal complementarity, with visuals, gestures, and speech perceived as working together more coherently.

\begin{figure*}[tb]
  \centering
  \includegraphics[width=\textwidth]{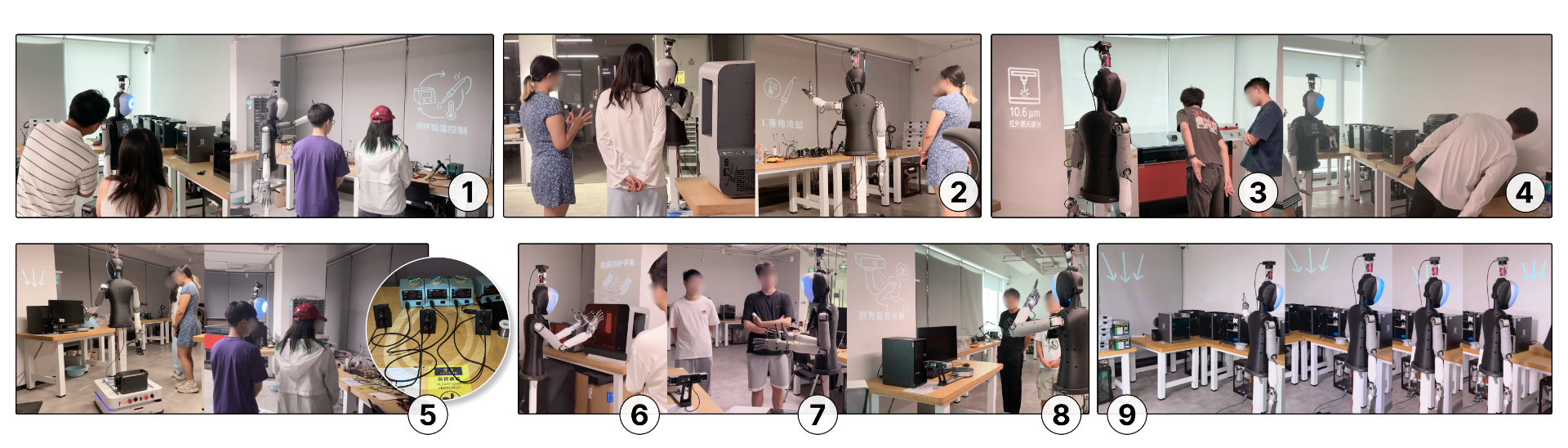}
  \caption{Illustration of qualitative findings with ProjecTA in real makerspace scenarios. Compared with the Baseline, ProjecTA \protect\circled{1} reduced attention switching between physical referents and displayed content, \protect\circled{2} enabled shared access to display information for the group, \protect\circled{3} enlarged critical information such as numeric data, \protect\circled{4} highlighted hard-to-see or hidden parts of equipment, \protect\circled{5} projected directional arrows and on-object highlights onto hand-indicated targets, \protect\circled{6} mimed protective glove use with a projected reminder, \protect\circled{7} used gestures to indicate size and range, \protect\circled{8} covered its eyes to introduce the "don't look at the laser" safety point, and \protect\circled{9} offloaded spatial descriptions by projecting arrows to indicate in-situ 3D printer locations.}
  \Description{A collection of nine numbered photographs showing the ProjecTA robot interacting with participants to illustrate qualitative findings.
  Panel 1: On the left, the participants look at the content shown on the robot’s chest-mounted screen. On the right, the robot projects a diagram of a soldering iron onto the wall for two participants, requiring minimal attention switching between the physical referent and the displayed content.
  Panel 2: On the left, one participant shifts position to see the screen display more clearly. On the right, the robot stands next to a machine, gesturing while projecting information that is visible to both participants.
  Panel 3: The robot projects critical numeric data onto a wall next to a laser cutter that two participants are observing.
  Panel 4: A participant bends over to inspect a 3D printer, guided by the robot.
  Panel 5: On the left, the robot uses gestures and projected directional arrows to indicate the 3D scanner. On the right, the robot combines gestures with projected highlights to point out the soldering equipment, with a circular inset providing a close-up view of the on-object highlights.
  Panel 6: The robot holds its hands in a protective posture while projecting an icon of gloved hands.
  Panel 7: The robot uses gestures to indicate size and range.
  Panel 8: The robot makes a gesture of covering its eyes with its hand.
  Panel 9: A composite image shows the robot projecting four sets of arrows sequentially onto a wall to indicate spatial locations.}
  \label{fig: Findings}
\end{figure*}

\subsection{Qualitative Results}
Qualitative results from participants' interviews help us further contextualize and exemplify their experiences:

\subsubsection{\textbf{How ProjecTA's In-Situ Projection Supported Learning in the Physical Space}}

Our results identified several key patterns in how the in-situ projection of ProjecTA brought unique advantages to participants' learning during the guided tour. 

\textbf{\textit{In-situ projection reduced attention switching between physical referents and displayed content.}}
As shown in \autoref{fig: Findings}~\circled{1}, by anchoring visuals adjacent to the target object, ProjecTA's in-situ projection reduced the learners' effort to switch between the visual display and the physical referents. Participants (23/24) consistently reported that placing cues on or near the objects reduced eye/head movements and shortened the path from locating the device to identifying the exact part. As P6 stated, \textit{``Projection placed the visual content next to the machine, so I can see the content besides the equipment at a glance, without switching back and forth''}. 
Similarly, P13 appreciated the projected content being \textit{``closer to the device,''} \textit{which ``naturally guides attention to the actual machine''} (P17). 
Several contrasted this with the chest-mounted screen: \textit{`` [with projection], I don't have to turn back to the screen; projection is more direct and makes it easier to engage with the target. With the screen, the constant attention switching makes it harder to focus''(P1)}. Additionally, Nearly half of the participants (11/24) mentioned that in-situ projection helped them feel less pressure of missing information. As P5 felt, \textit{`` turning from the screen to the machine risks missing an explanation''}. As P17 put it: \textit{``The screen feels discontinuous [...] when my attention switched from the screen to the equipment, [the Baseline] was still talking and showing visuals, and I felt I suddenly missed something''}. These accounts underscore in-situ projection's advantage over screen-centric paradigms in reducing misalignment and loss of information caused by gaze shifts. 

\textbf{\textit{In-situ projection improved shared access to information in the physical space.}}
In a group setting, the chest-mounted screen created a narrow, private viewing cone, hindering spatial access to the content (\autoref{fig: Findings}~\circled{2}). Participants reported having to reposition themselves to see the screen (P2-3), crowding and mutual occlusion when standing side-by-side (P7-8), or inconvenience to look at the screen from lateral viewing positions (P4, P7, and P22). 
Viewed through F-formation \cite{marquardt2012cross}, Baseline tended to fix the \textit{O-space} (primary interaction space) in front of its screen, forcing the participants to frequently shuffle and adjust head orientation to participate. As P3 noted: \textit{``During the laser cutter explanation, [the Baseline] was angled, so I had to [...] keep adjusting my position to catch the information''.} 
By contrast, in-situ projection created a large, public, and shared visual field that was legible from multiple angles. Participants noted: \textit{``you can see it clearly from all angles, and it's bigger'' (P2)},
and \textit{``even from the back I can still see it''(P5)}. 

\textbf{\textit{In-situ projection highlighted key information, and visualized hidden or hard-to-see parts.}}
In the physical learning setting, in-situ projection helped enlarge critical information and key components: 
P3, P7, P8, and P24 valued the usefulness of enlarging numeric data next to the machines, such as the working temperature of equipment, or the size of the workpiece.
As P3 stated, \textit{``the details such as numbers stand out more clearly''} (see \autoref{fig: Findings}~\circled{3}).
P7, P13, P14, and P18 had positive experiences regarding the projection using ``zoom-ins'' to help locate intricate components. 
P18 shared an example of this: \textit{``The projected [...] emergency STOP button immediately helped me locate it on the actual machine''} (see \autoref{fig: baseline_comparison}~\circled{4}). 
Moreover, participants appreciated that certain hard-to-see or hidden parts of equipment were `brought to the foreground' immediately by the projection. For instance, P13 shared an example of a back-side feeder reel (see \autoref{fig: Findings}~\circled{4}), which was hidden behind the machine. The projection helped them to notice and learn about this module. Similarly, P7 mentioned an example in which the projection helped them situate internal nozzles without direct sight.

\textbf{\textit{Challenges.}}
Despite its clear advantages, participants identified several practical challenges. Robotic arms' motion sometimes caused image jitter, which might be distracting (P16). Strong ambient lighting could diminish the projection's clarity and color fidelity(P5).
Finally, projecting onto uneven or angled surfaces introduced visual distortions. For instance, P18 reported and issue with perspective skew: \textit{``the projection is not a flat plane, there's a bit of oblique perspective.''}
P15 also reported an issue with occlusion, where a lower part of 
the projection became less readable when blocked by an object. 
Despite that the simplistic graphics were generally appreciated for their clarity, some participants (P3, P7, P8, P24) pointed out that when explaining rather complex parts, higher fidelity images such as photos of internal components might be more productive for learning.

\subsubsection{\textbf{Examples of Preferred Complementarity Across ProjecTA's Projection, Gesture, and Speech}}

Our results offered rich and vivid examples of how ProjecTA's projection, gestures, and speech supplement or amplify one another.

\textbf{\textit{Projection augmented or disambiguated deictic gestures for clear reference.}}
Upon experiencing Baseline, some participants indicated that the robot's deictic gestures (e.g., pointing) alone can be ambiguous sometimes: For instance, as felt by P15, \textit{``[Baseline] sometimes pointed not exactly at the content. I could infer its intent, but it didn't pinpoint the target''}.
In contrast, augmenting these deictic gestures with in-situ projections, such as directional arrows and on-object highlights, effectively resolved this ambiguity, enabling a smoother experience, especially when handing off instructions between devices (see \autoref{fig: Findings}~\circled{5}): 
\textit{``when switching to a new device, ProjecTA first projected an arrow onto the machine, followed by the pointing gesture, providing a more natural visual handoff''(P2)}. 
In cluttered environments, \textit{``a projected halo on the target surface plus a brief pointing motion helped me identify the object of explanation'' (P5)}. 

\textbf{\textit{Iconic and metaphoric gestures concretized or reinforced spoken and displayed explanations.}}
Over half of participants (15/24) recalled The robot's use of iconic and metaphoric gestures that demonstrated and reinforced the spoken or displayed content. 
They reported some memorable examples about its \textit{iconic gestures}, which directly mapped to physical actions or properties. For example, P4 described how ProjecTA \textit{``traced the distance with its hands at the 3D scanner to indicate size and scan range, synchronized with projected numbers.''}(see \autoref{fig: Findings}~\circled{7})
Other telling examples include miming a horizontal insertion (\autoref{fig: baseline_comparison}~\circled{1}) and then a subsequent operation, which aligned with the verbal and projected instruction (P22), or miming putting on protective gloves along with a projected reminder (P19 and P24, refer \autoref{fig: Findings}~\circled{6}), or performing a powder-mixing motion at the nylon printer (P10).
The robot's \textit{Metaphoric gestures} were experienced to symbolically reinforce abstract concepts communicated by speech and visuals at the same time.
Memorable examples included conveyance of prohibitions and warnings, such as the robot covering eyes to communicate `don't look at the laser' (P7, P9, P12, refer \autoref{fig: Findings}~\circled{8}), or using pushing motions or repetitive waving (\autoref{fig:teaser}~\circled{7}) to signal a warning (17/24). 
These embodied gestures created a strong link between the robot's physical motions and the knowledge to convey, enhancing the vividness and expressiveness of the spoken and displayed explanation.

\textbf{\textit{Projection offloaded spatial description from speech and clarified verbal content.}} Apart from disambiguating deictic gestures, projections was also experienced by the participants to offload and clarify the robot's verbal narration which otherwise might be complex or cumbersome.
Namely, with the projected cues illustrating the topic of discussion, the robot could replace verbose descriptions (e.g., `the second device from the left') with simple, direct phrases like `this device' or `this part' (see \autoref{fig: Findings}~\circled{9}). As P1 noted, \textit{``the projection made the part [of the scanner being talked about] explicit, so I knew exactly what part the ProjecTA meant.''} 
Beyond offloading references to concrete targets from verbalization, projections could also simplify the robot's verbal reference and description of abstract concepts: such as mentioning `this process' or `this phenomenon' verbally, while projecting an animated diagram or flowchart at the same time. 
This helped users grasp complex ideas more intuitively and seamlessly. 
For instance, a P1 experienced, \textit{``when some knowledge was explained [verbally], the visuals of projection helped me grasp how it works.''} 
Similarly, P14 found the projection offloaded retention, stating, \textit{``I don't always remember what ProjecTA says, but once I see the projection I remember it immediately.''}

\textbf{\textit{Challenges.}}
Despite the above mentioned benefits of such cross-modal coordination, the participants also pinpointed some pragmatic challenges for future refinement.
First, several (P11, P12) reported difficulty switching their focus between the robot's projections and gestures, noting that the two modalities, occasionally, competed for their attention rather than complementing each other.
Second, a few participants (P19, P23) reported that the robot's pointing sometimes seemed imprecise or skewed, causing them to narrow their focus to the projection alone.
Finally, several participants (P12, P17, P18) suggested adding even more verbal pre-announcement of the robot's intentions for upcoming visual or gestural outputs. This verbal cue would help them mentally prepare for the upcoming action, creating a smoother and more predictable interaction.

\subsubsection{\textbf{Social Expectations and Initiatives from Participants}}

Participants also articulated social expectations for a robotic TA to support the nomadic learning tour. Several participants (10/24) framed ProjecTA as a social coordinator that mediated group access to content, rather than just delivering information. For instance, P8 noted that \textit{``With the screen [of Baseline] I worried about bumping others; when two of us checked closely together it got crowded around the screen [...] whereas in-situ projection felt more naturally shared [...]''}. Echoing this, P16 noted that ProjecTA was better suited for facilitating \textit{``group access and joint attention''}, making it \textit{``easier for the group to orient to the same part of the equipment.''} 
Moreover, participants perceived the combination of projection and gestures as more socially coherent than the screen-gesture pairing; as P8 put, it \textit{``creates a sense of [...] inviting you to engage with the physical machine.''} By contrast, the screen-based condition: \textit{``the gesture is the gesture and the screen is the screen [...]''}

More than half of the participants (13/24) also reported that in-situ projections served as a trigger for their self-initiated exploration. P13, for example, recounted noticing a back-side feeder reel (see \autoref{fig: Findings}~\circled{4}) that had been hidden behind the machine until the projection highlighted it, which prompted him to step in toward the machine, inspect the referenced module, and verify the robot’s explanation. P7 further noted that \textit{ ``projecting the internal workings felt safer and more efficient for learning.''} These accounts suggest that projected content can create ``hooks'' that encourage learners to move, look closer, and personally verify robots' explanation, rather than only passively receiving information.

Many Participants (17/24) preferred the robot to communicate in human-like and emotionally engaging ways. For instance, P6 particularly appreciated the robot's welcoming poses at the start of the guiding tour and everyday metaphors in its explanations that helped them form intuitive understandings of technical components. P7 wished ProjecTA could adopt more expressive gestures and projected visuals for mobilizing learners’ emotions.

A few participants (3/24) also voiced a desire for more active, two-way interaction: they wanted not only to receive explanations but also to ask the robot questions during the tour. Although this feature was disabled in our study to maintain experimental control, this expectation indicates mixed-initiative dialogue as a natural extension. Finally, as ProjecTA moved between stations, P2 suggested that it could project its walking path and planned moves in the physical space to further increase its intuitiveness to learners in nomadic learning contexts.

\section{Discussion}
Robotic TAs can handle standardized, repetitive tasks and bring value to educational settings \cite{belpaeme2018social,rosanda2019robot}. Yet current deployments typically rely on a screen-based display \cite{tanaka2015pepper,yoshino2023teaching}, which makes it hard to move visual information with the narration and place it directly near or right on the object \cite{ahn2011projector}. Exploration on integrating in-situ projection with gesture-capable robotic TAs, as well as on aligning robots' projected visuals, speech, and gestures, remains scarce.
We thereby set out to explore how a robotic TA with in-situ projection, compared with a screen-based counterpart, affects learners’ experiences during makerspace tours (\textbf{RQ}).

To address this, we built ProjecTA, a robotic TA that links in-situ projection with speech and gestures to overlay information directly on or near target equipment for supporting learners in guided tours. In a real makerspace, 24 participants experienced projecTA and Baseline, its screen-based counterpart, in a controlled within-subject comparison.

Addressing \textbf{our research question (RQ)}, we used a mixed-methods approach. The quantitative results show that ProjecTA significantly lowered extraneous load with a large effect size, and enhanced perceived usability, usefulness of the visual display, and multimodal complementarity, indicating its ability to lessen unnecessary cognitive effort to improve learner experiences related to information delivery.
We then supplemented these results with qualitative analysis to characterize more vivid experiences: how near-object visual overlays aligned with robotic TAs' speech and gestures, reducing learners' referent matching and attention switching in the space, etc., which provided more contextual understandings to \textbf{RQ}.

To further extend our findings regarding our \textbf{RQ}, we distill design implications from our empirical data and discuss them below to inspire and inform future research. 

\subsection{Design Implications}

\subsubsection{\textbf{Implications 1: Unlocking More Design Possibilities Combining In-situ Projection with Deictic Gestures}} 

Our study shows that in-situ projection can serve as a means of spatial referencing and can be mutually reinforced by robots' deictic gestures. These findings meaningfully extend \textbf{DO1} summarized in our formative study, and point to three promising directions:

\textbf{\textit{Hand and projection co-referencing to spatial targets.}} Our study shows that projecting cues (arrows/halos) directly onto the hand-indicated target clarifies the reference, helping learners locate it quickly, reducing referent matching and gesture ambiguity. Building on this finding, and extending prior systems that implemented verbal-visual co-referencing \cite{narayanan2002multimedia,jiang2025visiobo} or projection-only referencing (e.g., supporting nursing \cite{bliss2022projected} and anatomy \cite{gao2021projector}), we envision hand–projection co-pointing where the robot points to the region with projected overlays marking detailed references such as paths, multiple sub-targets, or no-go zones, furthering current spatial referencing.

\textbf{\textit{Local cues for robot pointing at the projection.}} To reduce referential ambiguity when robots are pointing at projected content, we suggest adding local cues (e.g., arrows, halos, borders) on referenced targets, similar to ProjecTA pointing to physical targets. Beyond overlaying cues on real objects, 
we found that when the robot points to projected content, adding local highlights or animations can also be helpful, for reducing visual search and improving comprehension. Participants also explicitly expected the robotic TA to mark the exact region within the projected content it was indicating, so they could immediately see which part of the projection the gesture referred to, extending prior designs, which mainly focused on augmenting physical targets (such as Visiobo \cite{jiang2025visiobo}).

\textbf{\textit{Projection for hard-to-reach or untouchable referents.}} Many prior systems projected overlays to ease professionals' operational performance \cite{chakraborti2018projection}, while less work focused on helping novice learners grasp hidden, unsafe, or fragile parts of an object without touching it.
Building on our finding that in-situ projection revealed otherwise hidden details, we suggest that projection, as a precise, contact-free approach, could extend robot pointing for hard-to-reach or untouchable parts during nomadic learning tours (see \autoref{fig: choreography_workflow}~\circled{4}). For instance, one participant suggested projecting markers on an FDM printer's feed port and hot nozzle to avoid touch while speeding target localization. 
Beyond unsafe and risky scenarios, in our tours, projections onto hidden or hard-to-see parts prompted novices to move closer, inspect the equipment to check the robot’s explanation. These verification and exploration behaviors are consistent with prior educational psychology research showing that actively initiating self-explanation and testing one’s own interpretations can deepen conceptual understanding \cite{chi1994eliciting}.
More designs could draw from this beyond makerspaces, e.g., robots in a chemistry lab could indicate the storage area for concentrated sulfuric acid and project a warning sign on the bottle label, as inspired by AR safety training \cite{ismael2024acceptance}; In no-touch settings such as exhibitions, robot can indicate key details on the artifact without contact \cite{nikolakopoulou2022conveying}.

\subsubsection{\textbf{Implications 2: Thoughful Placement of Projected Content to Reduce Learners' Attention Switching}}

As some participants reported that the Baseline condition made them fear missing content and hesitate to look away from the screen, our qualitative findings in Section~6.2.1 similarly showed that ProjecTA’s near-object overlays reduced frequent referent matching and encouraged learners to verify information directly on the equipment.

Our study shows ProjecTA reduced learners' attention switching between presentation content and the physical equipment \textbf{(DO2)}. As shown in our qualitative results, some participants reported that Baseline made them fear missing content and hesitate to look away from the screen, whereas ProjecTA's near-object overlays reduced frequent referent matching. This extends prior findings in stationary tasks (e.g., assembly, repair, or driving \cite{tang2003comparative,aschenbrenner2019comparing,rosenthal2010augmenting,kim2009simulated}), suggesting the importance of thoughtful and finer-grained mechanism for the spatial placement of projected content in nomadic learning with mobile robots.

\textbf{\textit{Projectable regions of the artifact as the display}} Existing systems have extensively explored fixed-position projections, whereas our study surfaces design opportunities and challenges of designing mobile projective systems for learners.  
In our design, some equipment itself served as the projection surface, for instance, the nylon printer's body (see \autoref{fig:teaser}~\circled{2}). This way, overlays are placed directly on the suitable surfaces of explained artifacts, which could further reduce referent matching compared with wall projections \cite{rehman2020comparative}. To fully leverage this potential, a robotic TA needs robust 3D perception plus on-the-fly geometric and radiometric compensation, as shown in Raskar et al.'s Shader Lamps \cite{raskar2001shader}, which calibrates projection to an object's shape and color on-the-spot.

\textbf{\textit{Towards volumetric F-formation: robot–learner–object proxemics in 3D.}} Projection placement can be further optimized by analyzing the robot–learner–object proxemics. Our work used a pragmatic guideline to define projection placements: an F-formation with projected visuals and referents in O-space, surrounded by the robot and learners in R-space \cite{marquardt2012cross}. As shown in \autoref{fig: Fformation}~\circled{1}, the standard F-formation diagram is a preliminary 2D simplification \cite{marquardt2012cross}; to extend this, a more sophisticated 3D F-formation model (see \autoref{fig: Fformation}~\circled{2}) can be established to view O- and R-space as volumes and take learners' visual field, displays' viewing cone, and physical occlusions into consideration, as can be supported by spatial reconstruction (e.g., RGB-D mapping/SLAM \cite{newcombe2011kinectfusion}). 

Such a volumetric F-formation modeling could help a robotic TA properly decide when to utilize screen-based display and when to utilize in-situ projection. For example, when the human-robot formation is face-to-face dialogue or when artifacts are not suitable to project on, screens may be favored. 
Our ProjecTA hardware (see \autoref{fig: hardware}) was also built to incorporate both the chest-screen for frontal interaction and the in-situ projection for object-focused guidance.
In future real-world implementation, instead of treating projection and screens as competing options, designers can integrate both according to the 3D formation of learners and objects, to maximize the benefits of both.

\subsubsection{\textbf{Implications 3: Extending In-situ Projection as Human-Robot Collaborative Interfaces for Nomadic Learning}}

Different from whiteboards or fixed projectors, a robotic TA can carry its own projection. Our work illustrated how this could augment nomadic, walk-and-talk learning by visually supplementing the robot's verbalized content \textbf{(DO3)} and revealing critical or otherwise unseen information of the referent \textbf{(DO4)}. These position in-situ projection as a new communicative medium for robotic TAs and learners, pointing its evolution into learner-robot collaborative interfaces.

\textbf{\textit{Visualizing robots' chain of thought for learners' sensemaking.}} As shown in our results, in addition to equipment explanations, a few participants also wanted projected cues about ProjecTA's movement path and action plan, suggesting that enabling learners to predict a robotic TA's trajectories and status may increase the robot's social competence. Although our current work mainly focused on conveying equipment explanations, in future collaborative learning among robots and learners, in-situ projection could also externalize a robot's intent, plan, and intermediate reasoning to improve transparency and explainability \cite{schott2023literature}. For instance, Wengefeld et al. used laser projection to signify robot intent \cite{wengefeld2020laser}, and Mirror Eyes displayed mirror reflections on the robot's eyes to convey its focus of interest to humans \cite{kruger2025mirror}.
As educational research shows exposing AI reasoning can aid learning \cite{blasco2024impact}, future work could use spatial projection to depict robots' reasoning about and in the physical environments (e.g., spatial problem-solving) to support learners' sensemaking.

\textbf{\textit{Projected interface for learners' input.}} While our study primarily used in-situ projection as information display, projection can also function as an input interface, enabling spatial, embodied collaboration among learners and robotic TAs. Drawing from prior work, for instance, in material- and tool-based skill training, learners could trigger robots' support or feedback via on-projection inputs such as taps, traces, or region selection \cite{harrison2011omnitouch,xiao2013worldkit}. While prior studies already showed the value of a separate projector in turning its surrounding areas into interactive surfaces (e.g., \cite{ludwig2019printer}), our findings suggest future opportunities of such interactive surfaces being carried around by humanoid robots, serving as pervasive and environment-adaptive user interfaces to manipulate or collaborate with robots.

\subsubsection{\textbf{Implications 4: Fine-Graining Multimodal Orchestration for Projection-enhanced Robotic TAs}}
Our results show that the visuals, gestures, and speech were perceived to complement one and another significantly better in the projection-based condition than the screen-based counterpart.
Qualitative evidences further illustrate examples of such crossmocal complementarity preferred by the learners. Taken together, we outline the following design possibilities.

\textbf{\textit{Pre-action speech cues to better prepare learners for upcoming gestures and visuals.}} A few participants reported that projecTA's gestures and projections were occasionally unanticipated, primarily during their initial engagement with the system. 
A brief spoken cue in advance, such as ``Please see illustration above,'' prepares learners for robots' upcoming gestures or visuals. 
Such pre-action speech cues are especially helpful for collaborative tasks requiring frequent repositioning or viewpoint shifts, because gestures and projections are limited to each learner's field-of-view, whereas speech can broadcast across the space, prompting learners to anticipate upcoming visual communication.
Furthermore, as shown in our results, robotic TAs should pair such pre-action cues with brief pauses that not only give learners time to prepare, but also enable them to interrupt the robot, either requesting for more details, verifying what the robot has just explained, or changing the robot's upcoming actions. This helps learners to know when they can act without missing key information, and intervene the robot when needed.

\textbf{\textit{More precise temporal alignment in multimodal orchestration.}} 
While our system's temporal alignment technique proved effective, our results point toward the need for a more granular approach. Future work should pursue a finer-grained alignment by breaking down each modality into smaller, timestamped units. Further, individual keywords within speech segments, as well as keyframes within projected animations or videos could all be assigned with timestamps, to enable keyword-to-keyframe level visual-speech coordination.
Similarly, gestural units could be decomposed into timestamped micro-steps (e.g., `raise hand,' then `cover eyes,' then `lower hand,' instead of a single `cover-eyes' action). This aligns with the idea of `atomic actions' in LaMI \cite{wang2024lami}.
By aligning these multimodal micro-units based on semantic and contextual reasoning, we can achieve smoother and more expressive presentation of robots without sacrificing action transparency.

Additionally, our qualitative findings also revealed a few moments where concurrent modalities competed for learners' attention. This suggests that multimodal orchestration should not only synchronize channels, but also designate at each moment which modality leads and which ones recede into a supporting role. For instance, when an iconic gesture is intended to carry the main semantic load, projected content could briefly dim or simplify, or prompt learners to look back at the robot. Conversely, when detailed projected information should become the primary focus, the robot might momentarily pause unnecessary body movements and point toward the display. Such deliberate role-switching between lead and supporting modalities can reduce unnecessary modality conflicts and leverage the complementary strengths of speech, gesture, and projection.

\subsection{Limitations and Future Work}
To enable a controlled comparison, both ProjecTA and the Baseline delivered pre-choreographed explanations and did not support in-tour Q\&A or dialogue. 
Future implementation could introduce mixed-initiative interaction for more natural collaboration. 
The Presentation Choreography Workflow is promising to evolve into an educator-facing authoring tool for easy creation, customization, and fine-tuning. 
ProjecTA is an exploratory prototype; occasional image jitter was observed during motion. 
Future iteration could add a stabilizer or motion-compensated gimbal to the projector mount. 
Although the substantial reduction in extraneous cognitive load demonstrated ProjecTA’s promise for nomadic learning, immediate quiz scores remained comparable across conditions, suggesting the need for more comprehensive assessments of learning outcomes in future research.
Due to our focus on university makerspace, our sample consisted of novice makers, mostly from engineering or science backgrounds.
Future work can extend this line of inquiry to similarly complex and varied nomadic learning contexts beyond makerspaces (e.g., museums, botanical gardens, and even outdoor sites). It can also engage larger and more varied learner cohorts. These extensions will help better understand and generalize the impacts of similar systems on both learning performance and experience.
To focus on non-face-to-face, nomadic learning, the robot did not use facial expressions; in future, projecting facial cues onto the scene may provide social signals without prompting head turns. 
Finally, this study compared projection and screen presentation separately; combining body-mounted screens with in-situ projection may yield additive effects worth future evaluating.

\section{Conclusion}
In this study, we empirically examine how a robotic TA with in-situ projection, compared to a screen-based counterpart, affect learners’ experiences during makerspace tours. We implemented ProjecTA and Baseline that differed only in display modality. We evaluated them in a real university makerspace with 24 novices across two rounds covering six pieces of equipment. Results show that ProjectA significantly lowered learner's extraneous cognitive load and was rated significantly higher in perceived usability relative to the Baseline. Participants consistently reported that near-object visual overlays reduced attention switching, facilitated easier mapping from visuals to physical targets, and enhanced the multimodal integration of projection, gestures, and speech. Based on these findings, we distill design implications to inform future work in better supporting nomadic learning in physical settings with robotic TAs equipped with in-situ projection.



\begin{acks}
We appreciate all participating domain experts for their contributions to the two co-design activities, with special thanks to Prof. Fang Wan. Our thanks also go to all participants for their time and efforts throughout the process and the reviewers for their invaluable comments that helped us significantly improve this paper. This work is supported by the SUSTech Grant for AI: R01656020.
\end{acks}

\bibliographystyle{ACM-Reference-Format}
\bibliography{PJRobot}

\appendix
 \section{Semi-structured Interview Questions}
 \label{Semi-structured}
 \begin{itemize}
   \item \textbf{Noticing and visibility.} During the session, to what extent did you notice information presented via the \textit{projection} and the \textit{screen}? In what situations, if any, was the content difficult to see (e.g., viewing angle, glare, distance)?

   \item \textbf{Experiences with both display modes.} Looking back on both rounds, how would you characterize your interaction with the robot under each mode? Which aspects worked well or left a strong impression, which did not, and what concrete improvements would you suggest?

   \item \textbf{Comparative evaluation.} Comparing \textit{projection + speech/gestures} with \textit{screen + speech/gestures}, how did the two modes differ in supporting your understanding and attention? Please illustrate with a specific moment (e.g., when gesture–projection coordination worked or failed—what gesture occurred and what was shown), and indicate which mode you would prefer and why.

   \item \textbf{Potential application scenarios.} In what other scenarios do you think this system would be valuable? How do you envision it operating there, and what benefits would it provide?
 \end{itemize}

 \section{Familiarity and Past Engagement Rating Scale Results}
 \label{Appendix: familiarity}

 \begin{figure*}[tb]
   \centering
   \includegraphics[width=\textwidth]{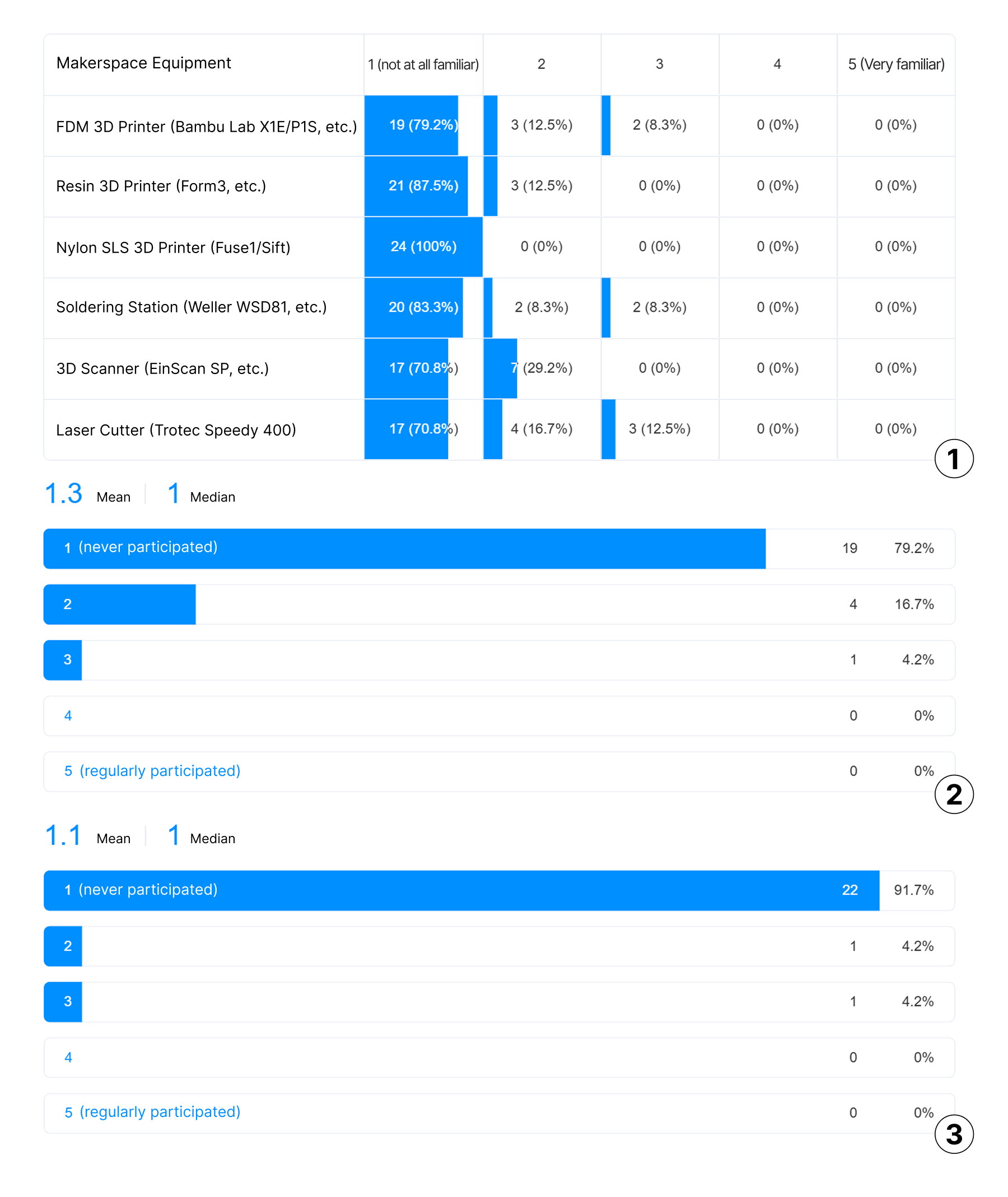}
   \caption{Demographic questionnaire results on \protect\circled{1} participants’ familiarity with makerspace equipment, \protect\circled{2} past engagement in hands-on activities, and \protect\circled{3} prior experience with equipment-guided tours.}
   \Description{A figure presenting three numbered tables of demographic survey results.
   Table 1 displays participant familiarity with six different types of makerspace equipment. Each equipment type is a row, and the columns represent a five-point familiarity scale from "1 (not at all familiar)" to "5 (Very familiar)." Data is shown as counts and percentages, with blue horizontal bars visualizing the distribution, which is heavily skewed towards "not at all familiar."
   Table 2 displays results for past engagement in hands-on activities on a five-point scale, from "1 (never participated)" to "5 (regularly participated)." Each row corresponds to a rating, showing the count and percentage of participants for that rating, also visualized with blue bars.
   Table 3 has the same structure as Table 2 but displays results for prior experience with equipment-guided tours. In both Table 2 and 3, the responses are predominantly concentrated in the lowest rating category.}
   \label{fig: demographic familiarity}
 \end{figure*}

\end{document}